\documentclass[usegraphicx]{mn2e}

\title[High redshift quasars]{High redshift quasars and the supermassive black hole mass budget: 
constraints on quasar formation models}

\author[J.~M.~Bromley, R.~S.~Somerville and A.~C.~Fabian]
{J.~M.~Bromley,$^1$ R.~S.~Somerville$^{2}$ and A.~C.~Fabian$^1$\\
$^1$ Institute of Astronomy, University of Cambridge, Madingley Rd., Cambridge
CB3 0HA\\
$^2$ Space Telescope Science Institute, 3700 San Martin Dr., Baltimore, MD 21218}

\begin{document}
\maketitle

\begin{abstract}
We investigate the constraints on models of supermassive black hole (SMBH) and quasar
formation obtainable from two recent observational developments: the
discovery of luminous quasars at $z\sim6$, and estimates of the local
mass density of SMBHs. If $\sim90$ per cent of this
mass was accreted at redshifts $z\la 3$, as suggested by the observed
quasar luminosity functions, these joint constraints pose a challenge
for models, which must account for the observed luminous quasar
population at $z\sim 6$ within a very limited `mass budget'. We
investigate a class of models based within the hierarchical structure
formation scenario, in which major mergers lead to black hole
formation and fuelling, and the resulting quasars shine at their
Eddington-limited rate until their fuel is exhausted. We show that the
simplest such model, in which a constant fraction of the gas within
the halo is accreted in each major merger, cannot satisfy both
constraints simultaneously.  When this model is normalized to
reproduce the number density of luminous quasars at $z\sim6$, the mass
budget is grossly exceeded due to an overabundance of lower mass SMBHs. We
explore a range of modifications to the simple model designed to
overcome this problem. We show that both constraints can be satisfied
if the gas accretion fraction scales as a function of the halo virial
velocity. Similar scalings have been proposed in order to reproduce
the local $M_{\bullet}-\sigma$ relation. Successful models can also be
constructed by restricting the formation of seed black holes to
redshifts above $z_{\rm crit} \sim 11.5$ or to haloes above a velocity threshold
$v_{\rm crit} \sim 55 \; {\rm km} \, {\rm s}^{-1}$, or assuming that only a fraction of major mergers
result in formation of a seed SMBH.  We also briefly discuss the issue of trying to assume a `universal $M_{\bullet}-\sigma$ relation' within the framework of simple Press--Schechter models, and further show that a fixed universal relation between SMBH mass and host halo mass is unlikely.
\end{abstract}

\begin{keywords}
accretion -- galaxies: interactions -- galaxies: active -- quasars: general
\end{keywords}

\section{Introduction}

In recent times, a growing wealth of observations has revealed a
strong correlation between the mass of supermassive black holes
(SMBHs) and the luminosity of the stellar bulges in their host
galaxies, and an even stronger correlation between SMBH mass and the
bulge velocity dispersion (Magorrian et al.\ 1998; van der Marel 1999;
Gebhardt et al.\ 2000; Ferrarese \& Merritt 2000; Tremaine et al.\
2002). This has in turn produced increasingly accurate estimates of
the overall local SMBH mass density.  If these SMBHs are, as now
widely believed, the relics of earlier quasar activity (e.g.\ Rees
1984 and references therein) then these estimates can place strong
constraints on the nature and evolution of the quasar population as
they effectively limit the `mass budget' available to account for
quasar activity.

The last few years have also seen dramatic progress in the
numbers of observed quasars as well as in the range of luminosities
and redshifts probed. The 2dF Quasar redshift survey (2QZ; Boyle et
al.\ 2000) and the earlier Large Bright Quasar Survey (LBQS; Hewett, Foltz \&
Chaffee 1995) have identified over 23000 quasars at redshifts $0.3 \la
z \la 3.0$. The Sloan Digital Sky Survey (SDSS; Gunn et al.\ 1998) has
observed over 650 confirmed quasars in the redshift range $3.0 \la z
< 5.7$ (Schneider et al.\ 2003) and a small number at $z\ga 5.7$ which
have been sufficient to constrain for the first time the density of
bright quasars at $z\sim6$ (Fan et al.\ 2001a, 2003). These objects
must host SMBHs with masses on the order of several $10^9\;{\rm
M}_{\sun}$ (Fan et al.\ 2001a, 2003; Willott, McLure \& Jarvis 2003).
Moreover, evidence has been found even in these highest redshift
objects for molecular gas (Bertoldi et al.\ 2003a; Walter et al.\
2003), dust (Bertoldi et al.\ 2003b), and metals (Freudling, Corbin
\& Korista 2003), underlining the fact that even at these early epochs
quasars are associated with galaxies developed enough to have
experienced significant star formation. 

In addition to the optical surveys already mentioned, observations of quasars and other active galactic nuclei (AGN) in
other wavebands have also been steadily improving.  This includes
substantial work done determining the soft X-ray quasar/AGN luminosity
function (Miyaji, Hasinger \& Schmidt 2000, 2001) and also several
imaging surveys of AGN selected in the hard X-ray band (e.g.\
Mushotzky et al.\ 2000; Giacconi et al.\ 2001; Hasinger et al.\ 2001;
Alexander et al.\ 2003; Barger et al.\ 2003) where samples are
unbiased by any line-of-sight absorption which may be present.  Thus
through multiwaveband observations progress is being made in piecing
together the history of AGN and their SMBH.

It has long been established that quasars, and indeed all forms of
AGN, are the result of gaseous matter
accreting on to a SMBH at the centre of a galaxy (Zel'dovich \&
Novikov 1964; Salpeter 1964; Lynden-Bell 1969; Bardeen 1970) and hence
must somehow be linked to galaxy formation.  Since that time a great
number of models have been published exploring quasar formation and
evolution within the context of the modern Cold Dark Matter (CDM)
paradigm of structure formation (including e.g.\ Efstathiou \& Rees
1988; Haehnelt \& Rees 1993; Haehnelt, Natarajan \& Rees 1998; Cattaneo, Haehnelt \& Rees 1999;
Kauffmann \& Haehnelt 2000; Cavaliere \& Vittorini 2000; Monaco,
Salucci \& Danese 2000; Nulsen \& Fabian 2000; Haiman \& Loeb 2001;
Wyithe \& Loeb 2002a, 2003; Hatziminaoglou et al.\ 2003; Granato et al.\ 2003; for a brief
overview of several of these models see Haehnelt 2003). However, none
of the above works have concentrated on successfully reproducing the
highest redshift population specifically within the constraints of the
limited `mass budget' implied by the estimated local SMBH mass
density. 

Any potential model of quasar formation finds itself faced with
three major unknowns. Firstly: where, how and with what mass do the initial
`seed' BHs form? Secondly: what events trigger their subsequent fuelling and growth, how much fuel do they supply and how efficiently is it converted into radiative energy?  Thirdly: how does feedback (from
star formation or the AGN activity itself) regulate and possibly even check their growth? These
uncertainties translate inevitably to a relatively large number of
free parameters in models.  In this Paper we aim to reduce this number by concentrating on the high redshift quasar population, allowing us to concentrate on just the first two issues: formation and fuelling.  In order to investigate
very luminous quasars such as those discovered at $z\sim6$, a very
large volume must be simulated so as to capture these extremely
rare objects. For this reason, we rely, as in the majority of the work
cited above, on the analytic Press--Schechter (Press \& Schechter 1974)
and extended Press--Schechter (Lacey \& Cole 1993) formalism to predict
the merger history of dark matter haloes. We consider a class of
models based on the basic assumption that seed black holes are created
in major mergers, and that these events trigger an episode of gas
accretion in galaxies with a pre-existing SMBH. We further assume that
SMBHs accrete and radiate at their Eddington-limited rate until their
fuel supply is exhausted, and that a fixed fraction of the accreted
mass is converted to radiation.

To rigorously explore this problem would require detailed modelling of
the star formation, gas cooling and feedback processes which alter and
compete for the intergalactic gas thought to fuel quasars. In this
Paper, we avoid these complications by focusing on constraints at very
high redshift $z\sim6$, when cooling times are short and only a small
fraction of the baryons is locked up in stars, so these competing
processes should be relatively unimportant. As well, we do not attempt
to model the complex interconnection of galaxy and BH/quasar formation
through feedback. Instead, we consider a suite of models with simple
parametrized recipes for SMBH formation and fuelling, which point the
way for more sophisticated and physical modelling. As we discuss in
the next section, attempts to combine data for observed AGN in X-ray
and optical wavebands now appear to be able to account for the
observed present day SMBH mass density, and moreover seem to suggest
that most of it was assembled at fairly low redshifts $z\la 3$.  This
enables us to place rough limits on the `mass budget' available at
higher redshifts. Combining this with the SDSS observations of the quasar
number density at $z \sim 6$ allows us to identify successful models
within the broad class we have considered, and to rule out other model
variants. 

Our results will in some sense reflect a `best case scenario' as
inclusion of star formation and feedback would presumably only act to
make fuel less plentiful for quasars, and thus strengthen our constraints.  As a result however it is
difficult to place strong constraints on the lower redshift behaviour
of the models as the characteristic upper peak and subsequent decline
of the quasar population number density at low redshifts is likely to be
strongly related to fuel depletion due to star formation (see e.g.\
Kauffmann \& Haehnelt 2000; Di Matteo et al.\ 2003).  Nevertheless we can demand that the
bright quasar population we track in our models never decreases in
number below that of the observed population at lower redshifts. In
practice this constitutes a relatively weak constraint. 

In Section~\ref{sec:dens} we detail our arguments limiting the SMBH
mass density at high redshift and present the criteria we shall use to
judge which of our quasar models are successful. In
Section~\ref{sec:framework} we present the basis for our family of
quasar models and the methods of our implementation.  We then present
in Section~\ref{sec:simple} a realization of the simplest member of
this family of models.  We show why such a model is not workable and
use it to demonstrate the challenges a successful model must face
before going on to explore other models within the family in
Section~\ref{sec:main}. In Section~\ref{sec:others} we briefly compare
our work to other recent results in the literature, and discuss the
implications of imposing a fixed scaling between black hole mass and
halo virial mass or velocity at all redshifts. We summarize and
conclude in Section~\ref{sec:conc}. Throughout this paper we assume a
`standard' $\Lambda$CDM\ cosmology with matter density $\Omega_0=0.3$,
vacuum density $\Omega_{\Lambda}=0.7$, Hubble parameter $H_0=70 \;
\rm{km}\, \rm{s}^{-1}\, \rm{Mpc}^{-1}$, and normalization $\sigma_8=1$, although our conclusions are not significantly altered for a lower value of $\sigma_8=0.9$ (the predicted high redshift SMBH mass densities would typically change by 5--10 per cent for most of the models considered here, and in no cases would change so much as to alter our conclusions as to the `success' of a model).

\section{Observational constraints}
\label{sec:dens}

We shall make use of two primary constraints in evaluating the models
considered in this Paper: the number density of luminous quasars at $z\sim
6$ and an upper limit on the total mass density of
SMBHs at $z\sim6$ (the `mass budget'). We will also consider a lower limit on the quasar number density
at lower redshift $2 \la z \la 6$ as a weaker constraint. In this
section, we walk through the arguments we used to obtain these
quantities from results in the literature. We also describe several
necessary conversions in detail.

The observed quasars at redshifts of $z \sim 6$ lie at the very limits of current detection and as such represent only the very brightest tail of the total high redshift quasar population.  Thus our models too will be concerned mostly with these brightest, rarest objects.  Our best information about the highest redshift quasar population
comes from a sample of $z > 5.8$ quasars identified by Fan et al.\
(2001a, 2003) based on \emph{i}-\emph{z} colour selection from Sloan
Digital Sky Survey (SDSS) imaging.  The sample consists of 6 quasars
in the redshift range $5.7 < z < 6.6$ and is complete down to an
absolute magnitude of $M_{1450}<-27.1$ (AB system; Oke \& Gunn 1983;
Fukugita et al.\ 1996).  It has a mean redshift of $\bar{z} = 6.08$
and predicts the number density of bright ($M_{1450}<-27.1$) quasars
at this redshift to be $5.2 \pm 2.1 \times 10^{-10} \; {\rm Mpc}^{-3}$
(all these figures take into account adjustments for our assumed
cosmology).  To compare this with the results from our models we need
to relate the magnitude limit of the sample to a minimum bolometric
luminosity for the quasars.  To do this we start with a given
bolometric luminosity and convert this to a rest frame $2500 \, {\rm
\AA}$ monochromatic luminosity using the bolometric corrections of
Elvis et al.\ (1994).  Using the mean spectral template of the same
authors we then convert this to a luminosity at the rest frame
wavelength of $1450\,{\rm \AA}$ which allows us to create a simulated
monochromatic magnitude $M_{1450}$ (AB system) for our
models\footnote{In fact the SDSS sample absolute magnitudes are
originally calculated for the rest frame $1280\,{\rm \AA}$ and then
the authors make their own corrections to arrive at the $1450\,{\rm
\AA}$ values, however the effects of this correction on our own
calculations will be negligible compared to our inherent
uncertainties.}. We find
\begin{equation}
L_{\rm Bol}=(1.0\pm
0.075)\times 10^{36.66-0.4M_{1450,{\rm AB}}} \; {\rm erg}\,{\rm s}^{-1}
\end{equation}
and hence that the SDSS sample magnitude limit of $M_{1450}<-27.1$
corresponds to a minimum quasar luminosity of $L_{\rm Bol} = 3.03 \times 10^{47}
\; {\rm erg} \, {\rm s}^{-1}$ in our models.

To determine the number density of quasars as bright as the SDSS $z\sim6$
sample at lower redshift, we use the luminosity functions determined
from the 2QZ and LBQS by Boyle et al.\ (2000).  These results are
quoted in absolute \emph{B}-band Vega magnitudes so we first make use
of the Elvis et al.\ (1994) bolometric corrections to find,
\begin{equation}
L_{\rm Bol}=(1.0\pm 0.06)\times 10^{36.59-0.4M_{B, {\rm VEGA}}} \; {\rm
erg}\,{\rm s}^{-1}.
\end{equation}
where we have assumed an offset of 0.12 mag between the Vega and AB
magnitude systems for quasar-like spectra (Schmidt, Schneider \& Gunn
1995; this offset is much larger than that typically used for magnitude conversions of stellar sources because of the large difference between stellar and quasar spectra). This implies that our minimum luminosity of interest, $L_{\rm
Bol} = 3.03 \times 10^{47} \; {\rm erg} \, {\rm s}^{-1}$, corresponds
to a \emph{B}-band magnitude limit of $M_B \leq -27.2$. We then
integrate the best-fitting luminosity function\footnote{Note that the
$\alpha$ and $\beta$ parameters for the luminosity function
$\Phi(M_B,z)$ in Boyle et al.\ (2000) are both missing minus signs in
the form in which the article was originally published (B.\ Boyle,
private communication).} to obtain the number density of quasars above
this magnitude limit, which at a redshift of $z=2$ is equal to $2.86
\times 10^{-8} \; {\rm Mpc}^{-3}$ (where we have corrected for the
lower Hubble parameter of $H_0=50 \; \rm{km}\, \rm{s}^{-1}\,
\rm{Mpc}^{-1}$ used by Boyle et al.\ (2000) in their $\Lambda$CDM
fit).  

Both the low and high redshift quasar number densities are based on
results from optical surveys, and it is worth noting that as a result
there is always a danger that obscured sources have been omitted from
our tally. However there is mounting evidence that the very brightest
members of the quasar population, which we are concerned with here,
are preferentially unobscured (see e.g. Fabian 2003 and references
therein).  A further potential source of error is that the corrections
and templates we have used from Elvis et al.\ (1994) are based on a
sample of low redshift ($z \la 2$) quasars, and little is known about
how well these templates apply to high redshift objects, although an
investigation by Kuhn et al.\ (2001) suggests that at least the
rest frame optical and UV energy distributions change very little for
quasars out to redshifts as far as $z \sim 3 - 4$.

The final quantity we wish to determine is an estimate of, or at least
an upper limit on, the total mass density in SMBHs at $z\sim6$. The
local SMBH mass density may be estimated by two independent
methods. One method makes use of the $M_{\bullet}-\sigma$ relation
plus the observed galaxy luminosity/velocity function to directly sum
up the mass present in remnant SMBHs at the present day (Aller \&
Richstone 2002; Yu \& Tremaine 2002).  These estimates now seem to be
converging on a value around $\rho_{\bullet}(z=0)=3 - 5 \times
10^5\;{\rm M}_{\sun}\,{\rm Mpc}^{-3}$. Specifically, Yu \& Tremaine
(2002) find $\rho_{\bullet}(z=0)=2.9\pm 0.5\times 10^5\;{\rm
M}_{\sun}\,{\rm Mpc}^{-3}$ while an earlier calculation of Merritt \&
Ferrarese (2001a) finds $\rho_{\bullet}(z=0) \sim 4.6 \times 10^5\;{\rm
M}_{\sun}\,{\rm Mpc}^{-3}$.

In the other approach, first attempted by So\l tan (1982), the
cumulative mass that must have been accreted by SMBHs in order to produce
the observed quasar luminosity function is summed up. In the past, the
best available estimates of the SMBH mass density $\rho_{\bullet}$
from these two very different methods have not always been in good
agreement. Improved estimates of the optical quasar luminosity function
(e.g. Boyle et al 2000; Fan et al.\ 2001b) are now available, and
extrapolating to higher redshifts and fainter magnitudes has produced
a much more complete estimate of the total contribution from accretion
by unobscured AGN (e.g.\ Chokshi \& Turner 1992).  The most recent
calculations (Yu \& Tremaine 2002) find that the total contribution by
the present day is around $\rho_{\rm acc}(z=0)=1.89\times 10^5\;{\rm
M}_{\sun}\,{\rm Mpc}^{-3}$ (assuming accreted matter is converted to
radiation with an efficiency of around 10 per cent) and indicate that
around 90 per cent of this total is due to accretion at redshifts $z
\la 3$.

However account must also be taken of the obscured population of
quasars whose demographics we do not yet understand so well.  We do
know there are large numbers of obscured weak AGN such as Seyfert IIs,
which will also contribute to the true value of $\rho_{\rm acc}$
(e.g. Alexander et al.\ 2001; Brandt et al.\ 2001; Rosati et al.\
2002), and indeed are required to explain the X-ray background (Fabian
\& Iwasawa 1999).  In addition there may also be a significant
population of faint highly reddened quasars (Barkhouse \& Hall 2001;
Wilkes et al.\ 2002; Richards et al.\ 2003a), and there is also some
evidence for a small population of totally obscured, Type II quasars
(e.g.\ Crawford et al.\ 2002; Norman et al.\ 2002; Wilman et al.\
2003; Zakamska et al.\ 2003) although the contribution of the latter
to the SMBH mass density is probably negligible.  Attempts to include
the effects of obscured sources in calculations of $\rho_{\rm acc}$
using X-ray observations (e.g.\ Fabian \& Iwasawa 1999; Elvis,
Risaliti \& Zamorani 2002) or by combining multiwaveband observations
(e.g.\ Salucci et al.\ 1999; Barger et al.\ 2001) have in the past
resulted in values of $\rho_{\rm acc}(z=0)$ that can actually exceed
the estimates of the total local mass density $\rho_{\bullet}(z=0)$
unless implausibly high radiation efficiencies are assumed. 

However, this overprediction may in part have been due to incorrectly
assuming that obscured and unobscured AGN populations peak at the same
redshift (e.g.\ Barger et al.\ 2001; Cowie et al.\ 2003), and a
picture is now emerging in which the accretion from the combined AGN
populations would seem to exactly account for the direct estimates of
the mass in SMBH remnants at the present day (Fabian 2003).  The high
redshift evolution of the obscured AGN population is much less well
known than that of the unobscured sources (see e.g.\ Hasinger 2002).
However because the period $z=0-3$ represents around 84 per cent of
the time since the big bang, we might nevertheless expect that the
bulk of the mass density from obscured sources would be deposited in
the epoch $z \la 3$, as is true for the unobscured sources.  Indeed
the (already fairly high) estimate of $\rho_{\rm acc}(z=0)$ made by
Barger et al.\ (2001) from multiwaveband observations of
\emph{Chandra} sources only considers the contribution made by
accretion for $z \leq 3$, so if this it to be consistent with other
measurements there cannot be very much contribution from higher
redshifts.

It seems then that almost all of the observed local SMBH mass density
is accounted for by radiative quasar accretion, leaving very little
room for any quiescent accretion modes.  Moreover the vast majority of
this mass must be deposited at redshifts $z\la3$, meaning that only a
very small fraction of the present day SMBH mass density
$\rho_{\bullet}(z=0)$ can be in place at high redshifts, i.e.\
$\rho_{\bullet}(z\ga 6) \ll \rho_{\bullet}(z=0)$. If we take the
upper bound on the current estimates on the local SMBH mass density,
i.e.\ $\rho_{\bullet}(z=0) \sim 5 \times 10^5\;{\rm M}_{\sun}\,{\rm
Mpc}^{-3}$, and then as a first approximation assume that around 90
per cent of this comes from accretion events below a redshift of $z
\sim 6$ (certainly we know for the unobscured sources this would be
true even for $z < 3$) then we arrive at an upper limit on the high
redshift SMBH mass density of $\rho_{\bullet}(z \sim 6) \sim 5 \times
10^4\;{\rm M}_{\sun}\,{\rm Mpc}^{-3}$.

Thus in this paper we shall consider a model `successful' if it both
reproduces the SDSS $z \sim 6$ population of bright quasars with a
SMBH mass density $\rho_{\bullet}(z \sim 6) \leq 5 \times 10^4\;{\rm
M}_{\sun}\,{\rm Mpc}^{-3}$ and continues to grow this population of
bright quasars so that they do not drop below the values predicted by
the Boyle et al.\ (2000) luminosity functions at $z=2$.  Because of
the uncertainties involved and the somewhat \emph{ad hoc} approach we
have taken to reach our mass density limit we shall consider models
`marginally successful' if they satisfy these constraints subject to
the more relaxed condition $10^5\;{\rm M}_{\sun}\,{\rm Mpc}^{-3} \geq
\rho_{\bullet}(z \sim 6) > 5 \times 10^4\;{\rm M}_{\sun}\,{\rm
Mpc}^{-3}$.

\section{Model framework}
\label{sec:framework}

In the hierarchical picture of structure formation, based on the  CDM
model, structures build up progressively over time,
the smallest objects collapsing first and then gradually merging
together to create larger and larger structures.  The much less
abundant baryonic matter initially traces the dark matter. But whereas
the dark matter is dissipationless the baryons are able, through
shock-heating and subsequent cooling, to collapse down to form much
more tightly bound structures -- the first (proto)galaxies -- within
the potential well of the collapsed dark matter `haloes'.  These
galaxies are then brought together by the subsequent hierarchical
mergers of their dark matter haloes where they can form early groups
and clusters or perhaps merge together themselves.

We base our model on the premise that quasar fuelling is accomplished
through tidal stripping of angular momentum from galactic gas during
such major galaxy mergers.  This popular premise has substantial
observational and theoretical support (Negroponte \& White 1983;
Barnes \& Hernquist 1991, 1996; Bekki \& Noguchi 1994; Mihos \&
Hernquist 1994, 1996; Bahcall et al.\ 1997; McLure et al.\ 1999),
although the mechanism responsible for the final inflow of the gas
once it has arrived within the last $100\;{\rm pc}$ from the galactic
centre is still unknown (for an alternative to the merger hypothesis see e.g.\ Kawakatu \& Umemura 2002; Granato et al.\ 2003).  In order to track the mergers of galaxies
within our model we construct Monte Carlo realizations of dark matter
`merger trees' following the algorithm of Somerville \& Kolatt
(1999). This algorithm allows us to reconstruct a typical evolutionary
history for a given final dark matter halo (the `root' halo) and
follow the hierarchical build up of its mass over time through
accretion and merger of smaller haloes as described by the extended
Press--Schechter probability distribution (Lacey \& Cole 1993). For a
root halo of given mass and redshift this process is averaged over a
large number of different realizations, allowing us to calculate the
average contribution (and its time dependence) made by the progenitor
haloes to all quantities of interest.  The simulations are repeated
over a grid of different root halo masses and the results are combined
by weighting them according to the probability of finding a halo of
that mass at the chosen output redshift.

We track the history of each halo back in time until we find a
progenitor halo with a circular velocity below $30\;{\rm km}\,{\rm
s}^{-1}$ or a mass below $1.7 \times 10^8\;{\rm M}_{\sun}$; the latter
mass corresponds to a circular velocity of $30\;{\rm km}\,{\rm
s}^{-1}$ at a redshift of $z\sim 20$, so we effectively track all
structures with circular velocities as small as $30\;{\rm km}\,{\rm
s}^{-1}$ back to redshifts of $z\sim 20$ and only progressively larger
structures beyond.  Structures smaller than $30\;{\rm km}\,{\rm
s}^{-1}$ are unlikely to be able to accrete gas efficiently in the
presence of a photoionizing background (see e.g.\ Gnedin 2000), and
since the recent results from \emph{WMAP} indicate that reionization
is likely to have occurred as early as $z=14-20$ (Kogut et al.\ 2003)
our chosen resolution should be valid out to similar redshifts.

Once we have created a merger tree, we associate baryonic material
with each newly formed dark matter halo.  This is assumed to fall
towards the centre of the resulting potential well and reside there as
a self-supported gaseous (proto)galaxy.  The initial amount of
baryonic matter available to a pristine dark matter halo is set by the
universal baryon fraction (here we use $f_b=0.13$, corresponding to
$\Omega_b=0.019h^{-2}$, which is consistent with the observations of
Levshakov et al.\ 2002).  This ratio remains fairly constant
throughout the history of a halo since the only mechanism for baryon
loss in our models is the conversion of accreted matter to quasar
radiation, which is fairly negligible.  We defer the consideration of
processes that may compete with quasars for the gas in these galaxies
(e.g.\ star formation, feedback and consequent heating) to a
subsequent paper (Bromley, Somerville \& Fabian 2004, in preparation) so as to
minimize the number of parameters in our model -- although we note
that this means our results will only apply to high redshifts where
gas is plentiful.

Whether such galaxies subsequently merge when their respective dark
matter haloes merge will depend on the rate at which they can lose
their orbital energy via dynamical friction against the background
dark matter.  When tracking halo-halo mergers within the merging tree
we label the central galaxy of the most massive dark matter halo as
the new central galaxy of the system and all the other galaxies become
`satellites'.  We compute the dynamical friction time-scale for these
`satellite' galaxies using the approximation of Binney \& Tremaine
(1987), modified to account for non-circular orbits (Lacey \& Cole
1993); see Somerville \& Primack (1999) for details.  If this
time-scale is shorter than the time to the next `branch' of the tree
(i.e. the next halo-halo merger), then the satellite is merged with
the central galaxy.  Otherwise the satellite remains until the next
halo-halo merger, when its `merger clock' is reset and the calculation
is repeated.

Whenever two galaxies merge, any central black holes they contain are
also assumed to merge instantly.  In addition, if a `major' galaxy
merger occurs (defined here as a merger in which the satellite galaxy accounts for at least 30 per cent of the total system mass), then some fraction \(f_m\) of the gas
is stripped of its angular momentum and falls to the centre, where it
can fuel a central SMBH.  The quantity $f_m$ is left as a free
parameter of the model.  

Once accretion on to a particular black hole has been `turned on' by
such a galaxy-galaxy merger, we assume that the accretion proceeds at
the Eddington limit until the fuelling gas supplied by the merger is
exhausted. The luminosity of an accreting black hole $L$ is governed
by the rate at which accreting matter is supplied $\dot{m}_{\rm
fuel}$, and the efficiency with which this accreted matter is
converted to radiation -- typically quoted in terms of the fraction
$\varepsilon$ of its total rest mass energy that is liberated.  Thus
\(L=\varepsilon \dot{m}_{\rm fuel} c^2\), where $c$ is the speed of
light. That fraction of the accreted matter which is not converted to
radiation adds to the mass of the black hole, and so the black hole
growth rate \(\dot{M}_{\bullet}\) is just
\(\dot{M}_{\bullet}=(1-\varepsilon)\dot{m}_{\rm fuel}\). Combining
these two expressions we find the quasar luminosity is related to the
black hole growth rate by \begin{equation}
L=\frac{\varepsilon}{1-\varepsilon}\dot{M}_{\bullet}c^2.
\end{equation} However, for Eddington limited accretion the black hole must be
shining at the Eddington luminosity where the radiation pressure
exactly balances the gravitational pull on infalling matter,
\begin{equation} L=1.15\frac{4\pi Gcm_p}{\sigma_e}M_{\bullet}
\end{equation} where \(m_p\) is the mass of the proton, \(\sigma_e\)
the Thompson cross-section of the electron, \(M_{\bullet}\) the
current mass of the black hole and the factor of \(1.15\) accounts for
the mean atomic weight per electron for a typical hydrogen and helium
gas mixture. Setting these two expressions equal yields an ordinary
differential equation with the solution \begin{equation}
M_{\bullet}(t)=M_{\bullet}(0){\rm e}^{t/\kappa}\end{equation} where
\(M_{\bullet}(0)\) is the initial black hole mass and the time-scale
\(\kappa\) is given by
\begin{equation}
\kappa=(1.15)^{-1} \frac{c\sigma_e}{4\pi
Gm_p}\frac{\varepsilon}{1-\varepsilon}.
\end{equation}

We can derive a similar expression for the amount of unaccreted gas
remaining $m_{\rm fuel}(t)$, and by setting \(m_{\rm fuel}(t_{\rm
Exh})=0\) find the amount of time \(t_{\rm Exh}\) for which a quasar
shines before all of the gas available to it is exhausted:
\begin{equation}
m_{\rm fuel}(t)= m_{\rm fuel}(0)-\frac{1}{1-\varepsilon}
\left[M_{\bullet}(t)-M_{\bullet}(0)\right]
\end{equation}
\begin{equation}
t_{\rm Exh}=\kappa\ln{\left[1+(1-\varepsilon)\frac{m_{\rm
fuel}(0)}{M_{\bullet}(0)} \right]}.
\end{equation}
For computational simplicity, we only record activity from accretion
on to central galaxies. Thus if a quasar is accreted into a new halo,
becoming a satellite galaxy, it will be `turned off' after a smaller
time-scale than the one given above. We find that this approximation
has a negligible effect on our results.

The exact value of the parameter $\varepsilon$ in all these equations
will depend on the nature of the accretion disc and the spin of the
black hole.  Taking into account photon capture by the black hole
itself, theory suggests $\varepsilon$ cannot exceed $0.057$ for a
static black hole and $0.4$ for a maximally rotating black hole,
although in fact accretion will act to bring the black hole spin to
its `canonical' value which reduces this maximum to $\varepsilon <
0.3$ (see Thorne 1974; in principle the effective efficiency could
exceed this limit since the disc luminosity could be augmented by
magnetohydrodynamic processes such as magnetic coupling with the SMBH, magnetic stress in the disc or torque on its inner boundary e.g.\ Gammie 1999; Krolik 2001; Li 2002; Wang, Lei
\& Ma 2003).  Most SMBH models assume the efficiency of a typical
accretion disc to be well below this maximum, at around $\varepsilon
\approx 0.1$, and we shall follow this convention, although in Section
\ref{sec:effic} we do discuss how the behaviour of our model is
affected by taking different values.

Clearly if one quasar differs from another only in that it shines
brightly enough to be visible for twice as long then it is twice as
likely to be observed.  Thus since we expect the bright lifetime of a
typical quasar to be much smaller than our model's intrinsic timesteps
(the times between halo and galaxy mergers) we group our results into redshift
bins of width ${\rm d}z=0.5$ and calculate the fraction of the time
spanned by the bin that each quasar spends shining above our magnitude
limit. When calculating number densities from our results we then
weight each quasar by this factor, in addition to the probabilistic
weight of its root halo discussed earlier.

The question remains as to how the initial `seed' black holes are
formed.  Several previous models have considered the possibility that
SMBHs form from much smaller ($\sim 10-100\; {\rm M}_{\sun}$) black holes,
remnants from either standard galactic star formation (e.g.\ Haiman \&
Loeb 2001) or from the very first stars -- the Population III stars
(e.g.\ Volonteri, Haardt \& Madau 2003).  However there are various
problems associated with such scenarios (see e.g.\ Haehnelt 2003).  In
particular some mechanism is required to facilitate their migration to
the centre of their host galaxy.  We instead focus on the possibility
that SMBHs form directly from the collapse of a large gaseous cloud in
the centre of a (proto)galaxy.  This was first discussed many years
ago (Rees 1984; Haehnelt \& Rees 1993; Silk \& Rees 1998); and
although it has obvious problems with the need to avoid fragmentation
during the collapse, it nevertheless remains a possibility and a
candidate mechanism has been sketched out by Haehnelt \& Rees
(1993). We shall further assume in our model that this collapse is
also connected to the major merger of galaxies -- that is, the same
tidally stripped gas which falls into the centre of a post-merger
galaxy and can fuel quasar accretion will also be viewed as the source
of the initial SMBH's formation. Precisely how \emph{effective} (how
likely it is to happen in a given major merger) this is and how much
of the gas this initial formation consumes (how \emph{efficient} it
is) will be left as free parameters of the model. We shall however
assume that the process is quiescent -- that is, that no radiation is
produced in this initial collapse -- although at high redshifts this
is unlikely to make much difference to our predictions as all but the
brightest of these `formation flashes' would be too faint to be
observable anyway.

\subsection{Model summary}

A brief summary of the steps involved in our modelling is as follows:
\begin{itemize}
\item Create hierarchical merger trees describing the formation history of dark matter haloes.
\item Associate baryons with each newly formed halo, forming (proto)galaxies.
\item At each level in the hierarchy, representing the merger of progenitor dark matter haloes, compute whether and when each of the constituent galaxies will merge (based on dynamical friction).
\item In every galaxy merger any existing SMBHs are merged.
\item In every major galaxy merger some fraction of the available baryons is assumed to be stripped of its angular momentum and fall to the centre, where it can either form a new `seed' SMBH or fuel an already existing SMBH.
\item Once provided with fuel a SMBH shines at the Eddington luminosity and accretes at the corresponding Eddington rate until this fuel is exhausted.
\end{itemize}
The mass ratio that constitutes a major merger is fixed at $0.3$ and
the radiation efficiency of accretion is fixed at $0.1$.  The
efficiency and effectiveness of SMBH formation, and the form of the
accreted fraction of baryons are left as free parameters.

\subsection{The radiative efficiency of quasars}
\label{sec:effic}

The efficiency of converting accreted material to radiation, given by
$\varepsilon$, is the controlling parameter in the equation for
Eddington limited black hole accretion.  A high value of $\varepsilon$
means that most of the accreting matter is turned to radiation, and so
only a small mass accretion rate is possible before the limiting
Eddington luminosity is achieved. Conversely, a lower value of
$\varepsilon$ means more mass must be accreted to produce the same
amount of energy in radiation and so the mass growth rate for a black
hole shining at the Eddington luminosity is higher.

The probability of observing an accreting black hole depends upon the
brightness of the accretion event and its duration.  If black holes
accrete at the Eddington luminosity, then their brightness is a
function only of their mass.  Thus the number of accreting black holes
above some luminosity depends on the number of black holes above a
particular mass limit -- which will depend on the rate at which black
holes have been able to grow prior to the epoch of observation --
while the duration of a given accretion event will depend inversely on
the black hole accretion rate if fuel is limited.  This means the
overall effect of the value of $\varepsilon$ on observations can be
complicated.

In our models we find that an increased $\varepsilon$ tends to
decrease the number density of quasars (above a given brightness
limit) at high redshifts, but increase the number density at lower
redshifts.  Thus $\varepsilon$ affects the steepness of the rise in
quasar numbers with time.

If fuel is plentiful then a lower efficiency model should grow faster.  However one would expect that once the rate at which major mergers can resupply a SMBH with fuel becomes the limiting factor then the number of bright quasars in a low and a high efficiency model would become equal.  In the slower accreting (high efficiency) model they would shine for longer however and so have a higher chance of being observed.  This indeed would seem to be the reason for the increased numbers of bright quasars at low redshift in our higher efficiency models.

However at high redshifts the situation becomes rather complex, and
there are three main factors which need to be take into account.
Firstly the equalizing between high and low efficiency models
discussed above may be less effective, since even though at redshifts
$z \sim 6$ the vast majority of quasar growth \emph{is} still limited
by the rate of major mergers, the time required to accrete the fuel
supplied just by a single major merger can in fact represent a very
large fraction of the total observed redshift bin at such redshifts.
Secondly it is the case that in a low efficiency model the total
amount of matter accreted by a SMBH (and hence its maximum luminosity)
will be slightly larger than in a high efficiency model since less of
the mass-energy of the fuel is converted to radiation.  While this is
a small effect, it nevertheless can make a difference when looking (as
we do) at the very brightest end of the quasar population.  Lastly an
additional complication arises if an accreting SMBH becomes a
satellite before exhausting its fuel and then merges with the SMBH of
a new central galaxy.  In this case the mass contribution will be
larger in the faster growing (low efficiency) models since the
satellite SMBH will have reached a larger mass.  These factors would
seem to combine to create the drop in high redshift quasars we see in
our higher efficiency models.

Indeed the last of these complications would seem to represent a possible barrier to accurately following high efficiency models within our current framework.  Although our standard choice of $\varepsilon = 0.1$ is hardly affected by the simplified treatment in our models whereby only accretion in central galaxies is followed, it would appear that once efficiencies as high as $\varepsilon = 0.2$ are considered SMBH growth becomes so slow that significant numbers of bright quasars become satellites before exhausting their fuel and so this simplifying assumption begins to break down.

Thus since a higher efficiency model both increases the low redshift quasar numbers, where our model is already poorly constrained, and also may need a more detailed treatment of satellite galaxies, we will delay further investigation of such models to a subsequent paper (Bromley et al.\ 2004, in preparation).

\subsection{Caveats and assumptions}
\label{sec:cav}

In this section we briefly discuss some caveats in our model and how
we have tried to deal with them, and also outline the possible
consequences of the assumptions we have made.  We begin by examining
some details in the algorithms of our simulation.

It is usual in Monte Carlo based merger tree models to weight each
root halo by the standard Press--Schechter distribution (Press \&
Schechter 1974) which provides the expected number density of haloes
of a given mass at a given epoch. However this distribution has
several known discrepancies when compared to N-body simulations (see
e.g.\ Somerville et al.\ 2000).  In an attempt to correct somewhat for
these, we use the corrected mass function of Sheth \& Tormen (1999) to
perform this weighting in our models (other `improved' mass functions,
e.g. Jenkins et al.\ (2001), would yield similar
results). Unfortunately, although a similar correction can be applied
to the conditional mass function used to generate the merger
histories, the resulting probability functions violate the Markov
condition and are no longer suitable for generating self-consistent
merger trees.
Therefore, to minimize the effects of such inaccuracies in the rest of
the merger tree, we resimulate each tree over a grid of output
redshifts, placing a root halo at each redshift bin we wish to record
data over (as usual averaging each over a grid of halo masses).

Similarly, on the topic of the merger tree algorithms, it should be
noted that due to the way our algorithm is implemented (see Somerville
\& Kolatt 1999 for details) it is sometimes the case that the very
earliest progenitor on a given `branch' of the merger tree may have a
circular velocity below our resolution limit.  Since our choice of
resolution was based on the notion that smaller haloes would be unable
to acquire baryons in a photoionizing background, we strip the
progenitor of all its baryons in such cases so as to remain
consistent with this premise.

Finally we note that since small mass ratio mergers are much more
common, the results of the model can be fairly sensitive to the exact
definition of a `major' merger.  In our model we have chosen to set
the boundary of the minor-major divide as the point where a merging
satellite galaxy accounts for 30 per cent of the total mass in the
system.  However we have found that all of the models we investigate
in this paper which prove successful for our chosen boundary can also
be made successful (by a suitable choice of the free parameters) for
values in the range 20 -- 40 per cent (although this is often at the
expense of steepening the rise in quasar numbers with time which could
have implications for the low redshift quasar population not
investigated here)\footnote{It should be noted that while it appears
usual in quasar models to consider the fraction of the total system
mass as the basis for the minor-major divide (as we do here), many 
galaxy formation models (which use the divide in determining starbursts) instead use the ratio of baryonic masses to
classify mergers as major vs. minor; e.g. Somerville \& Primack (1999);
Cole et al. (2000).}.

Concerning the assumptions made in our model, there are three main
points which need to be addressed.  Firstly we note that our
assumption that quasars always shine at the Eddington luminosity right
up to the point where their fuel is exhausted is somewhat unphysical.
Realistically a quasar's light curve is likely to rise steeply as it
is first activated, approach the Eddington luminosity and then decay
away as fuel becomes scarce; however the exact form such a curve would
take is not known and will affect the overall shape of the quasar
luminosity function (see e.g.\ discussion in Cattaneo
2001). Nevertheless the effects of a decaying tail in the quasar light
curve are likely to be most important in reproducing the faint end of
the luminosity function, and since we are concentrating in this paper
on the bright quasars visible at high redshift, the simple
approximation we use is probably sufficient.  Were we to include such
an effect then we suspect it would require if anything slightly more
numerous or more massive quasars to still account for the observed
high redshift quasar numbers, making the constraints from the SMBH
mass density presented in this paper even tighter.

Secondly we need to consider the effects of our formation and
accretion scenarios.  We have assumed that formation of an initial
seed SMBH requires a major merger, as does subsequent accretion.  A
consequence of this is that unless a halo acquires its SMBH from a
minor merger, it must experience at least two major mergers in order to activate a
quasar. Thus at high redshifts, when few SMBHs have been formed,
quasar activity will already be biased towards haloes that are in
`merger-rich' environments.  This is in contrast both to previous
models which have seeded haloes with black holes completely
independently of mergers, and to those which have not differentiated
between accretion and formation but treated them as the same process
(examples of both types of model are discussed in
Section~\ref{sec:others}).  When we examine model variants with a
reduced SMBH formation efficiency (i.e.\ where not all the gas freed
in the major merger is used in creating the initial SMBH), we have
continued in the same spirit by assuming that the gas not consumed by
the initial formation is expelled from the galaxy by some outflow
process (of course it may be that the fraction of gas able to reach
the galactic centre is slightly less anyway in the case where no prior
SMBH exists).  For reasonably large efficiencies this does not seem
too unlikely; most of the models we shall show to be successful have
initial SMBHs with mean masses in the range $\sim 5 \times 10^3 - 5
\times 10^5 \; {\rm M}_{\sun}$ (at $z=6$) which would indeed release a
substantial amount of gravitational binding energy upon formation and
could conceivably expel the remaining gas back out of the central
inner parsec of their galaxy.  However as the formation efficiency
becomes smaller this rapidly becomes less and less likely, so clearly
we should treat as suspect a model which relies on particularly small
efficiencies (which it would be hard to justify physically anyway).
At the end of Section~\ref{sec:results} we briefly discuss the effects
on the model if we remove this assumption and instead allow gas unused
in SMBH formation to accrete on to the newly formed SMBH in the next
timestep.  However, since our timesteps are the intervals between
galaxy and halo mergers they are not evenly spaced, this is only a
very crude treatment and we do not spend much time on it.  A better
approach would have been to allow accretion of leftover gas directly
after the formation; but this would require a sufficient understanding
of the formation process to be able to estimate the time-scales
involved and without the inclusion of star formation and other
processes that would compete for this gas the treatment is likely to
remain unrealistic anyway.

Lastly we examine the assumptions we have made to arrive at the
constraints used to judge our models.  We first consider the number
density of quasars at $z \sim 6$ calculated from the SDSS data.  While
we have assumed that the value found by Fan et al.\ (2003) represents
the true density of the population, our models are not in fact very
sensitive to changes in this value by a factor of a few.  Although
none of the six quasars in the SDSS $z \sim 6 $ sample show multiple
images (Fan et al.\ 2003; see also Richards et al.\ 2003b),
nevertheless the discovery of galaxies near to the line of sight of
two of these quasars suggests that two of the sample may in fact be
magnified by lensing (Shioya et al.\ 2002; White et al.\ 2003; Wyithe
2003).  If the intrinsic luminosity of these two quasars before
magnification is in fact below the SDSS magnitude cut, this would
reduce the number density of bright ($M_{1450} < -27.1$) quasars at $z
\sim 6$ by a factor of $\sim \frac{2}{3}$.  This figure is consistent
with an earlier study by Wyithe \& Loeb (2002b) which concluded
lensing of the SDSS sample is unlikely to have increased the true
number density at $z=6$ by much more than a factor of $\sim 1.5$.
Changes at this level would not however affect our models
significantly, and so we do not feel that the possible presence of
lensing will affect our conclusions.

In constraining our high redshift SMBH `mass budget' we have further
assumed that all the accreted mass from quasars combined with the mass
of their initial SMBHs is accounted for by the observed central SMBHs
in present day galaxies -- that is, that the evolution of SMBHs is
lossless.  However there are two mechanisms that might lead to mass
loss in the evolution of central galactic SMBHs. Firstly it is
possible that the merger of SMBHs is a lossy process, as potentially a
significant amount of the mass-energy of a binary SMBH system could be
released as gravitational radiation in the last stages of its merger
(see e.g.\ Yu \& Tremaine 2002).  Additionally, it is possible that
the time-scale for SMBH merging is sufficiently long that there is a high
probability of accreting a third SMBH, which is likely to lead to the
ejection of the smallest SMBH or in some cases even the ejection of all three
SMBHs via the 'slingshot' mechanism (see Haehnelt
\& Kauffmann 2002; Volonteri et al. 2003, and references therein). Any
such ejected SMBHs that escape agglomeration in the nucleus of their
host galaxy would not be included in the kind of census used to
estimate the local SMBH mass density, and thus effectively constitute
another mode of mass loss.  If either of these processes is significant then this would weaken our constraints on the high redshift `mass budget'.

\section{Model results}
\label{sec:models}

We now examine several different variants of the models within the
framework just described.  We begin with the simplest case, which
turns out to be unable to satisfy our joint constraints, and then add
additional parameters in an attempt to find one or more successful
variants.

\subsection{The simplest case}
\label{sec:simple}

To realize a particular model we need to specify the fraction of gas
accreted on to central SMBHs in major mergers, $f_m$ (which may be a
function of other variables), and the effectiveness and efficiency of
the formation of the initial seed SMBH.  Perhaps the simplest scenario
is to take $f_m$ to be equal to a fixed value and assume a completely
effective and efficient SMBH formation process -- that is, in every
major merger in which the remnant does not contain a SMBH, all
the tidally stripped gas falling to the centre is assumed to form a new SMBH. The single free parameter,
$f_m$, may then be fixed by requiring the model to reproduce the
observed number density of bright quasars at redshift $z\sim 6$. We
find that a value $f_m=0.0095$ achieves this.

\begin{figure*}
\includegraphics[width=0.5\textwidth]{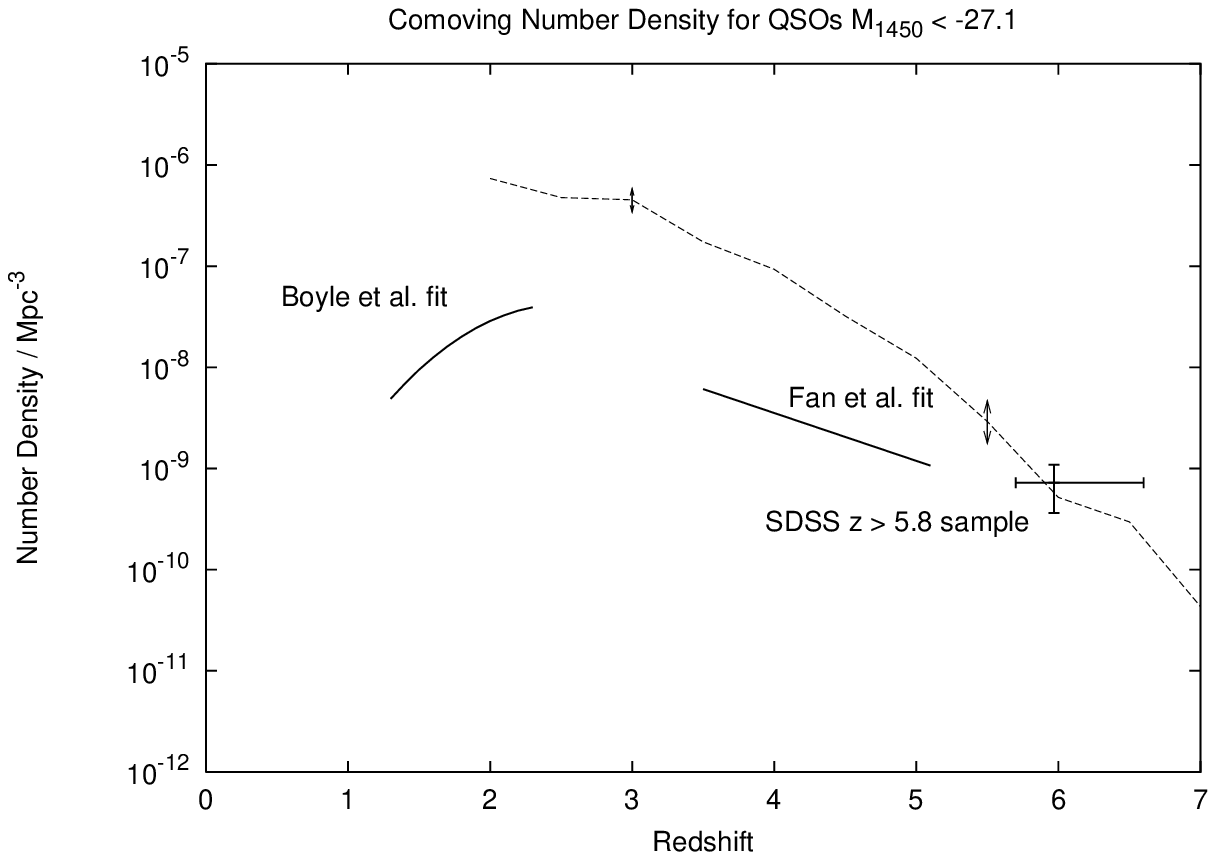}\includegraphics[width=0.5\textwidth]{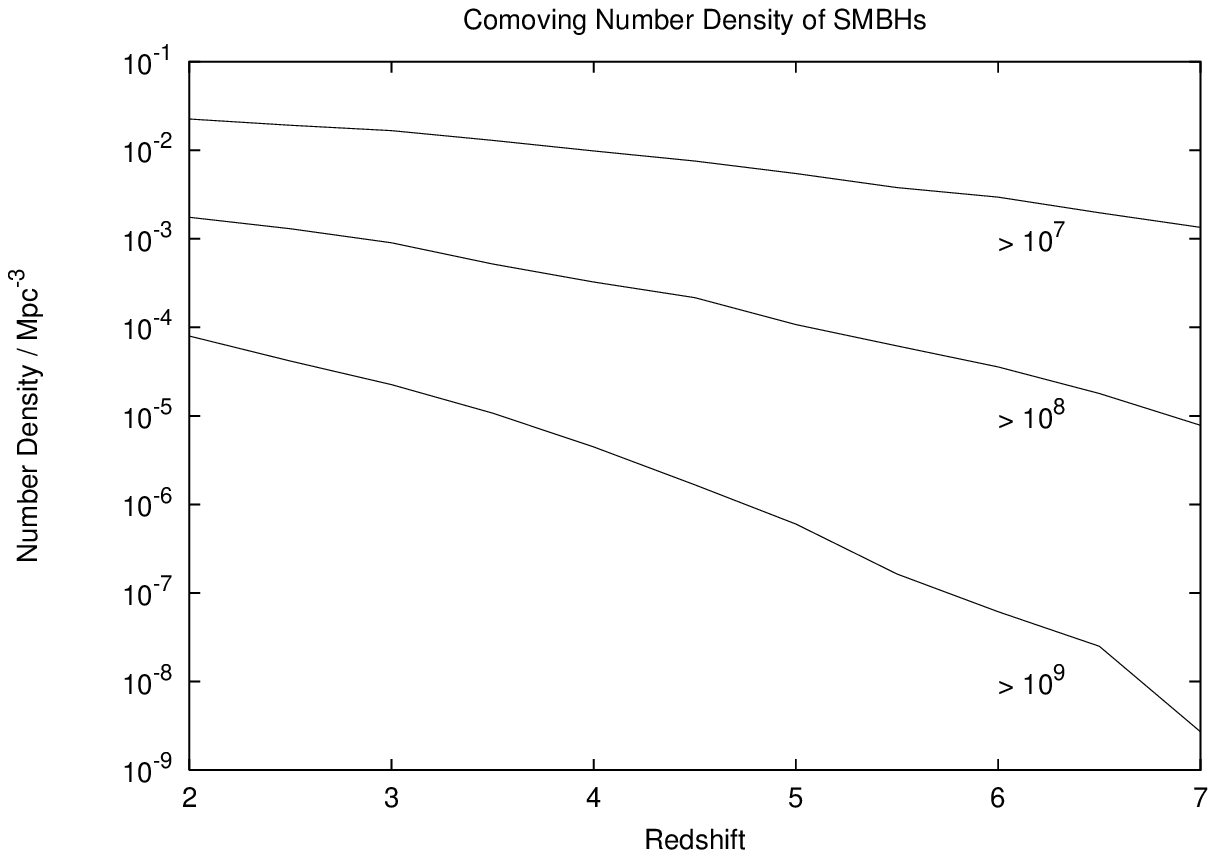} \\
\includegraphics[width=0.5\textwidth]{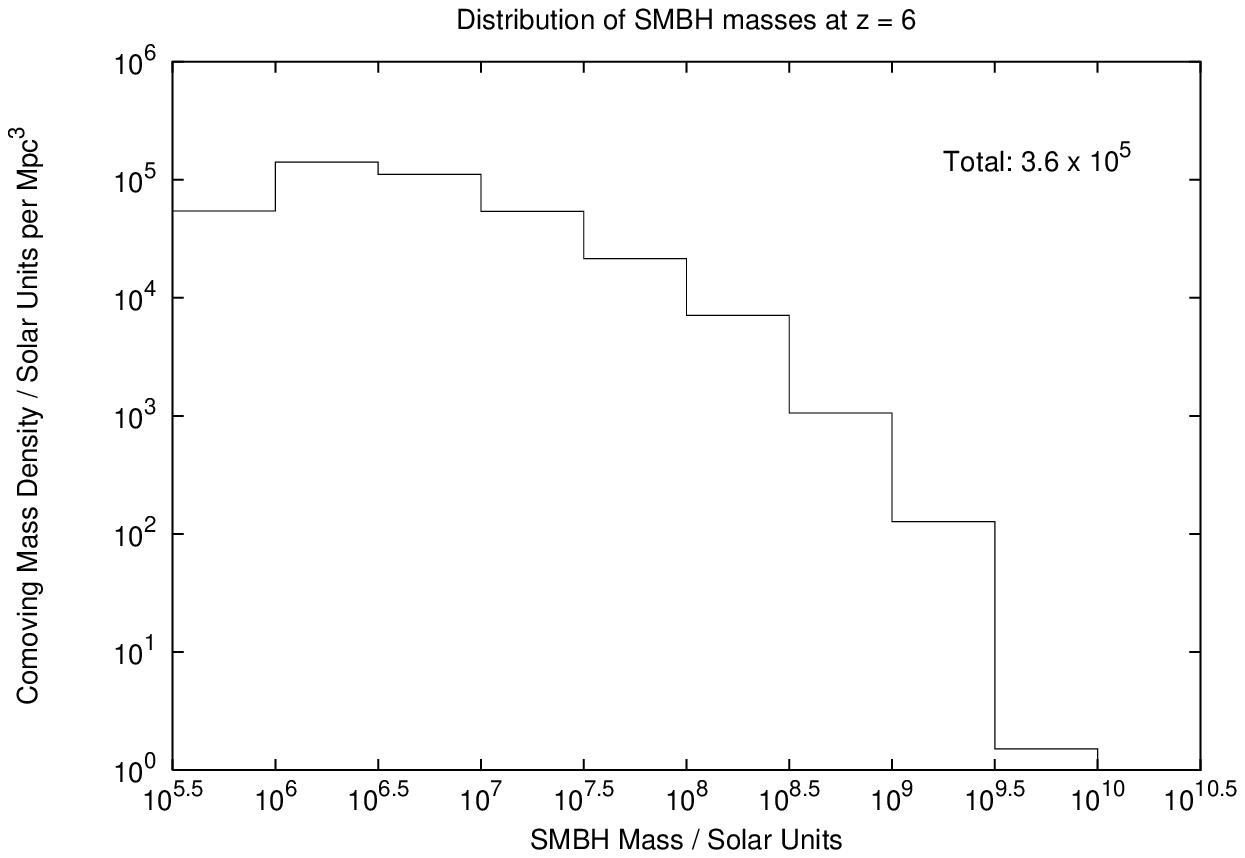}\includegraphics[width=0.5\textwidth]{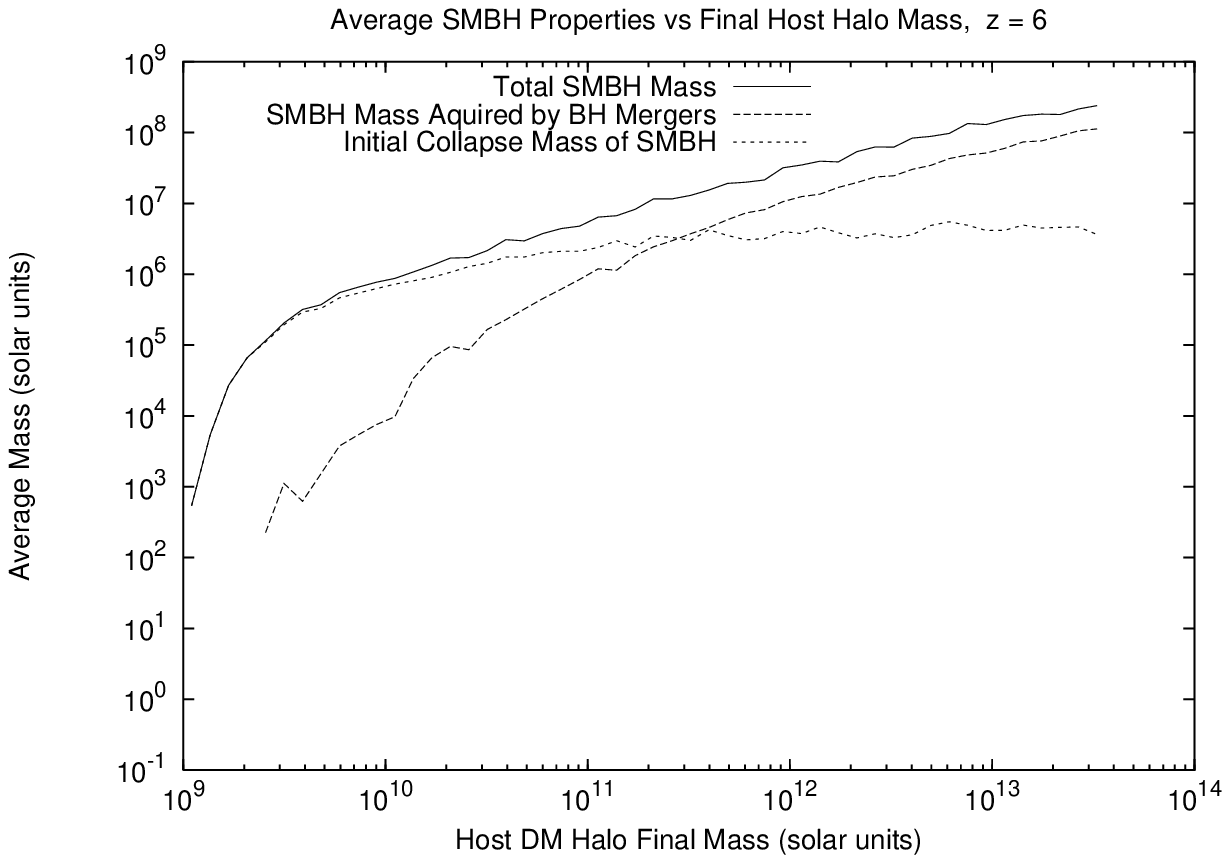}
\caption{Results from the simple model, with a constant accretion
fraction $f_m=0.0095$ and completely efficient and effective formation
of seed SMBHs. (a) \emph{Upper Left Panel:} The comoving number density
of bright ($M_{1450}<-27.1$) quasars (dashed line) as a function of
redshift, shown with the observational estimates (solid lines; see
text for details). (b) \emph{Upper Right Panel:} The cumulative
comoving number density of SMBHs larger than $10^7$, $10^8$ and
$10^9\;{\rm M}_{\sun}$ as a function of redshift. (c) \emph{Lower Left
Panel:} The mass distribution of SMBHs in terms of their contribution
to the total SMBH mass density at $z=6$. (d) \emph{Lower Right Panel:}
Average SMBH mass at $z=6$ as a function of host halo mass (solid) and
the contribution to this from mergers (long dashed) and initial
collapse (short dashed), the remaining contribution being from
accretion. }
\label{fid}
\end{figure*}

We show several results from this model in Figs.~\ref{fid}a--d.  In
the top panels we plot the number density of luminous
($M_{1450}<-27.1$) quasars and of SMBHs as a function of redshift.  In
Fig.~\ref{fid}a, the vertical arrows indicate the typical uncertainty
on the mean, calculated from both the scatter over different realizations of
halo merger histories and the errors in our luminosity binning due to
the uncertainties in the quasar bolometric correction.  Because it is
not clear that the sampling errors of different luminosity bins are
independent the errors are not combined in quadrature but added
directly (reflecting the worst case scenario).  Although the errors
are based on the assumption that the variance of the underlying
distribution can be approximated by the standard estimator for a
Gaussian, comparing several different runs of the model suggest they
provide a fairly good indication of the true errors.  In the bottom
panels we show the contribution to the mass density of SMBHs at $z=6$
as a function of SMBH mass, and the average mass of the central SMBH as a
function of host halo mass, also at $z=6$ and with an indication of how much of this mass is due to black hole mergers and formation (seeds).  The latter plot is created
by averaging the properties of the SMBH in final root haloes of
different masses at $z \sim 6$ (actually $z=5.75$) over a large number
of realizations (note that by construction the SMBH of a `root halo'
always corresponds to the SMBH in the central galaxy). 

We can see that, when normalized at $z=6$ as described above, the
model overpredicts the number density of luminous quasars $N_{\rm
QSO}$ at lower redshift, by a factor of $\sim 25$ by $z=2$. As we have
discussed, this is likely to be due to our neglect of star formation
and feedback. However, the more serious problem is that when this
model is normalized to reproduce $N_{\rm QSO}$ at $z=6$, it inevitably
and seriously overproduces the SMBH mass budget, already producing by
$z=6$ a total mass density of $3.6 \times 10^5 {\rm M}_{\sun}\,{\rm
Mpc}^{-3}$, in excess of the observationally derived value of Yu \&
Tremaine (2002) for $z=0$.

Part of the problem is that only a small fraction of SMBHs are
`activated' by major mergers at a given time. To be luminous enough
to make it into the SDSS $z\sim6$ sample, an Eddington limited quasar
must possess a SMBH of at least $M_{\bullet}=2\times 10^9\;{\rm
M}_{\sun}$. Only about 1/100 of the galaxies with such massive SMBHs are
active at a given time at such redshifts. However the main reason that
the model so badly exceeds the SMBH mass budget can be seen in
Fig.~\ref{fid}c: most of the total mass density is contributed by low mass
SMBHs, objects which even when accreting at the Eddington rate would not
be luminous enough to be detected by the SDSS at this redshift.

The short-dashed line in Fig.\ \ref{fid}d shows the mean initial mass for SMBHs (i.e.\ the mass they contain after their initial gaseous collapse) as a function of the mass of the dark matter halo the SMBH finally ends up in at $z \sim 6$.  Interestingly this appears to become almost flat as the host halo mass increases, although the mechanism for this is not obvious.  To a first approximation the initial SMBH mass will just be a function of the mass of the halo it first formed in.  Thus one possibility is that this flattening of the curve represents an upper limit to the likely mass of those haloes which form their own SMBHs,  larger haloes being very likely to simply acquire their SMBH from such smaller haloes which are incorporated into their collapse (due to the nature of hierarchical structure formation).  However this mean behaviour does not preclude the possibility of significant outliers, and it is another less apparent failure of this simple model that some of the
initial seed SMBHs have extremely large masses -- in some cases as large
as $\sim 10^9\; {\rm M}_{\sun}$. 

 While we are assuming that fragmentation
is somehow sufficiently suppressed to allow the initial SMBH to
collapse, nevertheless it seems unlikely that such a process could
continue to scale up to such large masses.  Possibly the formation process would in fact be rather inefficient in its use of gas.  Alternatively it may be that,
because we have neglected the star formation and feedback which
probably will consume much of the gas in the halo, the amount
of gas that is available at the first major merger is unrealistically
large.

A simple way to correct this problem would be to make the more
physical assumption that the collapse process will not be 100 per cent
efficient, but rather only some fraction of the total gas involved in
the collapse ($f_{\rm seed}$) will actually end up forming a SMBH. A
hard upper limit to the initial mass ($M_{\rm seed, max}$) could also
be imposed to cover the case of formation in extremely large early
galaxies.  We will assume the gas not used in the collapse is lost in an
outflow (or in any case does not become available for later
accretion), in this case one might hope that this correction would also help to
correct the other problem -- the excess total SMBH mass density. This
is because haloes which form SMBHs but then experience little or no
subsequent accretion will make a smaller contribution to the total
SMBH mass density. However in practice this is counterbalanced to a strong degree
by the need to increase $f_m$ in order to maintain the agreement with
the observed bright quasar number density at $z=6$, because the
 SMBHs responsible for these quasars must now be grown from smaller seeds.  As a result we find that with a
cap of $M_{\rm seed, max} = 10^6 M_{\sun}$, we must reduce the
formation efficiency to $f_{\rm seed} \sim 1.8 \times 10^{-4}$ in
order to reduce the SMBH mass density at $z\sim 6$ to our target
value (the model is `marginally successful' with $f_{\rm seed} \sim
0.036$).
Such a low efficiency is probably unlikely in any case (as discussed in Section~\ref{sec:cav}), but also the
required increase in $f_m$ is sufficiently large that the models even
more severely overpredict the number density of luminous quasars at
low luminosity (by a factor of $\sim 220$). It remains to be seen
whether the competing effects of star formation and feedback can
account for these discrepancies. For the moment we proceed under the
assumption that this can only be part of the solution.

\begin{figure}
\includegraphics[width=0.5\textwidth]{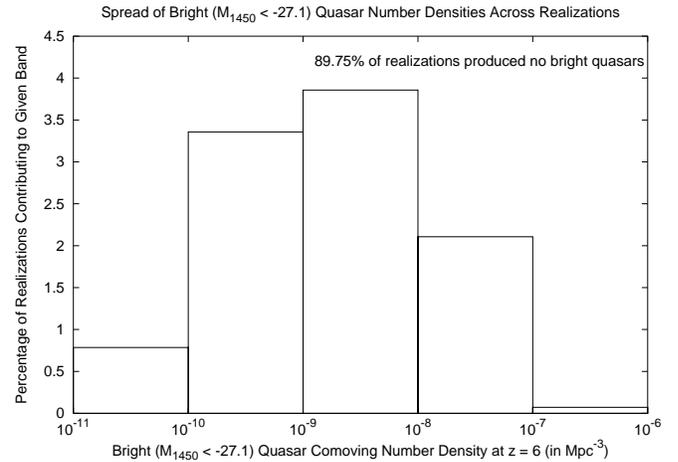} \\
\caption{The distribution of the number density of bright
(\(M_{1450}<-27.1\)) quasars at a redshift of \(z=6\) obtained in
individual realizations of the model, grouped in terms of the
percentage of realizations producing values within the range of a
given logarithmic bin.  The model uses the same parameters as that
shown in Fig.~\protect\ref{fid}, and once all the realizations are
averaged together matches the observed number density at $z=6$.  Note that only
$\sim10$ per cent of the realizations produced non-zero number
densities.  Had our simulation included rarer, larger mass haloes then the lower density bins would likely become more populated.  However the mean is dominated by the 2--4 per cent
of the realizations producing values in the range $10^{-7} - 10^{-9}$ (which would not be affected by considering larger haloes).  Even significant increases in the lower density bins from running the simulation over larger haloes would not change this (and so would not affect our results).
}
\label{numerics}
\end{figure}

It is worth mentioning that because quasars such as those found in the
SDSS $z > 5.8$ sample (which are scattered over 2870~deg$^2$) are so
intrinsically rare, we find we must average over around 700 Monte
Carlo realizations of our model before we get good convergence. For
example, a single realization of the model is the result of a
Press--Schechter weighted average of a grid of root haloes of 50
different masses, producing \(\sim 6.17 \times 10^5\) separate galaxy
merger events between redshifts \(z = 5.75 - 6.25\), of which on
average $\sim 26000$ are major and of these only \(\sim 380\) will
generate quasars with \(M_{1450}<-27.1\). In fact about 90 per cent of
the 700 realizations at $z=6$ did not produce a
single quasar above this magnitude limit. The distribution of the quasar
number densities found in the other $\sim10$ per cent of the
realizations is shown in Fig.~\ref{numerics}. 

\begin{figure}
\includegraphics[width=0.5\textwidth]{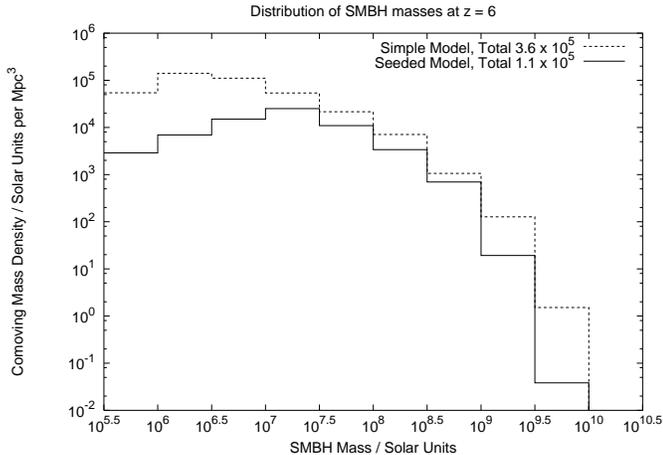} \\
\caption{Results for an alternative scenario where SMBHs grow from
$10\;{\rm M}_{\sun}$ seeds created once a host halo's virial velocity
exceeds $40 \; {\rm km}\, {\rm s}^{-1}$ (solid line) compared to our
initial simple model (dashed line). The plot shows the mass
distribution of SMBHs (grouped in logarithmic bins) in terms of their
contribution to the total SMBH mass density at $z=6$.}
\label{vcdemo}
\end{figure}

It is natural to ask whether the rather strong conclusions reached
above are specific to the particular approach we have taken to form
seed SMBH. Clearly when compared to `seeding' mechanisms which use small seeds, our
method would place more BH mass in galaxies which acquire an initial
black hole but then see little accretion. However, we have seen that
the end result was not very different when we reduced the mass of the
seeds, as more accretion was required to continue to match the
observed quasar number density at $z=6$. One could argue that for this
reason our result should not be very sensitive to the details of the
recipe for creating seed BHs.  To confirm this we have run our simple
model with an alternative scenario in which haloes are seeded with
$10\;{\rm M}_{\sun}$ BHs as soon as the virial velocity reaches some
critical value. The results are shown in Fig.~\ref{vcdemo} for $v_{\rm
crit} = 40 \; {\rm km}\, {\rm s}^{-1}$, illustrating that as long as
we normalize our model to $N_{\rm QSO}$ at $z=6$, the problem of
exceeding the SMBH mass budget remains.  This is true regardless of
the choice of the critical velocity. Although the total mass density
is indeed reduced, it still just exceeds the value that we deem
`marginally successful'. Also, as with the previous experiment in
which the efficiency of seed formation was lowered and the seed mass
capped, the number density of luminous quasars increases too rapidly,
overproducing the observed values at lower redshift much more severely than our original model (increasing the seed mass reduces this problem but exacerbates the overproduction of the SMBH mass density).  Thus, although we restrict our attention to the `gaseous
collapse' model for seed BH formation in the rest of this paper, it is
likely that our conclusions will also be pertinent to models considering other 
mechanisms.

\subsection{Model variants}
\label{sec:main}

We have shown that the model with the minimal number of free
parameters cannot satisfy our requirements. In this section we explore
other models with a minimal number of additional parameters. Our
simple model had two main failings: firstly, it overproduced the total
mass in SMBHs, depositing too much mass in small SMBHs; and secondly the masses of the seed SMBHs could be
excessively large.  As already mentioned, the latter can be easily
resolved by decreasing the efficiency with which seeds are formed
and/or capping the seed mass; the former however would seem to require
some way to skew the mass function of SMBHs at early times towards higher
masses. There are several approaches one could take to try and achieve
this.

One method would be to scale the fraction of gas accreted $f_m$ as a
function of the host halo properties.  Within our simple framework a
dark matter halo has only two defining properties: its virial velocity
and the redshift at which it collapsed (its mass and density are a
function of these two quantities). The two basic possibilities are
therefore to scale $f_m$ with either virial velocity or collapse
redshift.  This scaling could be in the form of some smooth functional
relation or could just be a simple truncation.

Alternatively we could consider the properties of the SMBH itself as a
basis for determining the value of $f_m$.  In this case the only
defining property is the mass of the SMBH. By varying the value of
$f_m$ with SMBH mass one could alter the relative importance of black
hole mergers and gas accretion according to the stage of evolution of
a given SMBH, concentrating the accretion of gas on to the more
developed, more massive SMBHs.

Another possibility is to more drastically modify the conditions for
seed SMBH formation.  If the probability of forming a SMBH in a given
major merger is made small, then galaxies in environments where
mergers are frequent will preferentially begin growing SMBHs earlier
than those in poorer environments, because they are more likely to
form them (through major mergers) or acquire them (through minor
mergers).  Thus early SMBH growth will be biased toward galaxies with more mergers, where SMBHs are more likely to be able to grow to large sizes.  Indeed it has been shown by Menou, Haiman \& Narayanan
(2001) that in principle only a small fraction of the galaxies at high
redshifts need to host SMBHs in order to account for the ubiquity of
SMBHs at the present epoch.  A similar effect could be achieved if rather than a probabilistic
approach, the host halo properties were used to determine when a
seed SMBH forms, since again if SMBHs are initially rare then they are more likely to be acquired by galaxies which experience more mergers.

While we clearly cannot hope to explore every possible variant of each
of these approaches, we will examine a set of 7 possible models --
labelled A through G -- which cover the main alternatives (summarized
together with the results of the next Section in Table
\ref{tab:summary}).

In the first three models we will examine various ways of scaling the
accreted gas fraction $f_m$.  In Model A we consider the case of a
redshift dependent scaling $f_m \propto (1+z)^n$ where we shall vary
the value of $n$.  In Model B we consider a scenario where the
accreted fraction scales with the square of the halo virial velocity
$f_m \propto v_{\rm vir}^2$, motivated by the fact that the binding
energy of the halo is expected to scale in this way.  In Model C we
adopt the functional form suggested by Kauffmann \& Haehnelt (2000),
\begin{equation}f_m \propto \frac{1}{1+(280\;{\rm km}\,{\rm
s}^{-1}/v_{\rm vir})^2}.\label{eqn:kh}\end{equation}

The SMBH mass itself is used as the governing parameter in Model D
where we keep a fixed $f_m$ but only allow accretion to take place
when SMBHs exceed a mass of $10^6\,{\rm M}_{\sun}$, allowing them
only to grow through merger with other SMBHs prior to this.  Loeb \&
Rasio (1994) have shown that stable accretion on to a SMBH at the
centre of a galaxy cannot take place if the SMBH is less massive than
$\sim 10^6\,{\rm M}_{\sun}$. While accretion driven by a major galaxy
merger is unlikely to be directly comparable to continuous stable
accretion there nevertheless may well be a minimum mass necessary for
the process to be effective and this is our motivation for considering
this model.

Finally, with the last three models we explore various approaches in
which the recipe governing the formation of seed SMBHs is altered.  Note
that in these models we do not allow stripped gas to accumulate in the
centre of galaxies when no SMBH is present.  In Model E we consider a
case in which seed SMBHs only form in major mergers occurring before some critical redshift $z_{\rm
crit}$.  Along similar lines, Model F hypothesizes that SMBHs can only be
formed and fuelled in haloes with virial velocity above some critical value $v_{\rm
crit}$.  Lastly Model G posits that the probability of forming a seed
SMBH in any given major merger ($P_{\rm seed}$) is less than unity. 

To prevent the formation of very massive seed SMBHs, for all of the
models we additionally place a hard upper limit on seed masses of
$M_{\rm seed, max} = 10^6\;{\rm M}_{\sun}$ and also reduce the seed SMBH
formation efficiency to $f_{\rm seed} = 0.1$.  For comparison we will
also show the results of the original simple model under these same
assumptions about seed SMBH formation, which we label Model S.

\subsection{Results}
\label{sec:results}

The models were run over a variety of different parameter values and
for a minimum of 700 realizations in each case.  The results are
summarized in Table~\ref{tab:summary} where we show the SMBH mass
density achieved at $z=6$ and the average initial SMBH mass at this
redshift. Each model is adjusted to fit the observed quasar number
density at $z \sim 6$, and assumes $f_{\rm seed}= 0.1$ and $M_{\rm
seed, max}=10^6\;{\rm M}_{\sun}$. Except where otherwise noted, all major
mergers produce seed SMBHs if no existing SMBH is present. We also
indicate any other important properties for each model, and in addition show the number density of bright quasars at $z=2$ predicted by the models.  While the latter will inevitably greatly exceed the observed values, due to the lack of star formation as discussed earlier, the relative values can be used as a rough guide to the relative strengths of star formation and feedback each model would require if it were to be extended to low redshift.
\begin{table*}
\begin{tabular}{|c|c|c|c|l|}
\hline Model & $\rho_{\bullet}(z=6)$ & $\bar{M}_{\bullet , {\rm init}}(z=6)$ & $N_{\rm QSO}(z=2)$ & Description / Comments \\
\hline S & $1.2 \times 10^5$ & $1.5 \times 10^5$ & $1.3 \times 10^{-6}$ &
Original simple model (seed SMBH formation occurs in all major mergers,\\ 
 & & & & $f_m$ = constant), with $f_{\rm seed}= 0.1$ and $M_{\rm seed, max}=10^6\;{\rm M}_{\sun}$. \\
 & & & & \\
A & $1.3 \times 10^5$ & $1.2 \times 10^5$ & $2.4 \times 10^{-7}$ &
$f_m \propto (1+z)^n$\\
& & & & \emph{Comments:} 
Less successful than original simple model (see text). \\
& & & & Results shown here for $n=1$. \\
 & & & & \\
B  & $4.7 \times 10^3$ & $4.0 \times 10^3$ & $9.5 \times 10^{-7}$ & $f_m \propto v_{\rm vir}^2$ \\
 & & & & \\
C  & $1.5 \times 10^4$ & $1.3 \times 10^4$ & $1.3 \times 10^{-6}$ &
$f_m \propto [1+(280\;{\rm km}\,{\rm s}^{-1}/v_{\rm vir})^2]^{-1}$ \\
 & & & & \\
D  & $4.8 \times 10^4$ & $1.6 \times 10^5$ & $1.5 \times 10^{-6}$ &
$f_m$ constant but accretion only allowed when $M_{\bullet} > 10^6 \; {\rm M}_{\sun}$. \\
& & & & \emph{Comments:} Maximum value of $f_{\rm seed}$ for (marginal) success is $0.1$ ($0.2$). \\
 & & & & \\
E  & $3.8 \times 10^4$ & -- & $1.2 \times 10^{-6}$ &
Seed SMBHs formed only at redshifts $z > z_{\rm crit}$; $f_m=$ constant. \\
 & & & & \emph{Comments:} Results shown for $z_{\rm crit}=12$.  For (marginal) success we require \\
 & & & & $z_{\rm crit} \geq 11.5$ ($8.0$). \\
 & & & & \\
F  & $7.9 \times 10^2$ & $10^6$ & $1.9 \times 10^{-6}$ &
 Seed SMBH formation \emph{and} accretion only take place in haloes with \\
 & & &  & $v_{\rm vir} > v_{\rm crit}$, $f_m$ is constant when accretion occurs. \\
 & & & & \emph{Comments:} Results shown for $v_{\rm crit} = 150\; {\rm km} \, {\rm s}^{-1}$.\\
& & & & For (marginal) success we require $v_{\rm crit} \geq 55\; {\rm km} \, {\rm s}^{-1}$ ($35\; {\rm km} \, {\rm s}^{-1}$).  \\
 & & & & \\
G  & $7.5 \times 10^3$ & $4.1 \times 10^5$ & $1.8 \times 10^{-6}$ & 
 Seed formation occurs with probability $P_{\rm seed} < 1$; $f_m$ is constant. \\
 & & & & \emph{Comments:}
 Shown for $P_{\rm seed} =0.02$,  (marginal) success requires \\
 & & & & $P_{\rm seed} < 0.3$ ($0.85$). \\
\hline
\end{tabular}

\caption{Summary of the model variants. For all models, $M_{\rm seed,
max} = 10^6\,{\rm M}_{\sun}$, $f_{\rm seed} = 0.1$ and the
normalization of $f_m$ is chosen so the models reproduce the number
density of observed quasars at $z \sim 6$.  If a model has further
additional parameters these are also specified. SMBH mass density
($\rho_{\bullet}$) is given in units of ${\rm M}_{\sun}\,{\rm
Mpc}^{-3}$ and the mean initial SMBH mass ($\bar{M}_{\bullet , {\rm
init}}$) in units of ${\rm M}_{\sun}$. Recall that our preferred upper
limit on the observed SMBH mass density at $z=6$ is $\rho_{\bullet}= 5
\times 10^4 \; {\rm M}_{\sun}\,{\rm Mpc}^{-3}$ and the value for
`marginal success' is $\rho_{\bullet}= 1 \times 10^5 \; {\rm
M}_{\sun}\,{\rm Mpc}^{-3}$. Also shown is the predicted number density of bright ($M_{1450}<-27.1$) quasars at low redshift, $N_{\rm QSO}(z=2)$, in units of ${\rm Mpc}^{-3}$, which as already discussed is expected to greatly exceed the observed value ($2.9 \times 10^{-8} \; {\rm Mpc}^{-3}$).}
\label{tab:summary}
\end{table*}

Model A, in which the fraction of accreted gas scales as $(1+z)^n$,
does not perform too well.  For $n>0$, accretion becomes rapidly suppressed at lower redshifts, and so the number of bright quasars increases much more
slowly with time. In some cases this model may even fail to reproduce
the quasar number density at $z\sim 2$; however the SMBH mass density at
$z\sim 6$ is still far too high.  As with our initial simple model
this can be rectified by taking a very low seed formation
efficiency, however the values required are even more extreme that for
the original model -- an efficiency of $\sim 9.5 \times 10^{-5}$ is
required when $n=1$ and $\sim 4.2 \times 10^{-5}$ when $n=2$ to match
the target value of the SMBH mass density.  This is presumably because
at very high redshifts where most of the seed SMBHs are formed the
accretion fraction is greatly enhanced.  

One might then expect that Model A would be more successful for
negative values of $n$ (so that the accretion fraction increases with
time), however this is only partially true.  Taking a value of $n=-1$
we do indeed find that a somewhat larger value of $f_{\rm seed} \sim 5
\times 10^{-4}$ now suffices to make the model successful, thanks to
the reduced accretion fraction at very high redshifts. However because
of the rapid increase of $f_m$ with time, we find that the
normalization required to match the data at $z \sim 6$ would probably
result in an overproduction of quasars at low redshift, since it means 38 per cent
of all the available gas would be accreted in mergers by $z=0$. Larger negative
values of $n$ naturally exacerbate this problem.

The possibility remains that some model where the accreted gas fraction rises
initially with time but then levels off to avoid the latter problems
might be successful.  However we suspect that in fact it will not be
possible to achieve significant gains with any model in which the
accreted gas fraction scales as a function of redshift only, since
fundamentally (as we saw in Section \ref{sec:simple}) we require the
scaling to reduce the contribution from haloes with comparatively
small black holes, most of which are likely to have formed at higher
redshifts.  But any overall decrease in the accretion fraction at these
high redshifts will make it very difficult to reproduce the SDSS
observed quasars as these are already such rare objects.  What seems
to be needed is an accretion fraction which varies according to
individual halo properties at each epoch.

Indeed, Model B, in which the fraction of gas accreted in mergers
scales as $v_{\rm vir}^2$ seems very successful.  The overall mass
density of SMBHs is dramatically reduced, and the mass density in the
most massive SMBHs ($10^9 \; {\rm M}_{\sun}$ and larger) is reduced by
a smaller amount comparatively, since the largest haloes where such
SMBHs are hosted get preferentially more accretion.  Furthermore the
most massive SMBHs gain a significantly larger fraction of their mass
from accretion (rather than from mergers or the initial collapse).
This latter effect is probably a combination of the fact that large
haloes can accrete a larger fraction of the available gas, as well as the fact that the SMBHs
contributed by mergers of smaller haloes will not have been able to
grow as much.
Model C, using the scaling from Eqn.~\ref{eqn:kh} for the accretion
fraction, has similar behaviour to Model B, but is not quite as
successful.

Model D, in which accretion can only take place on to SMBHs larger than
$10^6 \; {\rm M}_{\sun}$ also proves successful, although only just.
The total SMBH mass density is reduced by strongly suppressing the
growth of smaller SMBHs, since they are restricted to growing by direct BH-BH mergers.  Interestingly the mass density contributed by different mass SMBHs is lowest for SMBHs that are very near the transition mass of $10^6 \; {\rm M}_{\sun}$ in the model.  This is because SMBHs can only grow to this size by the slow process of SMBH mergers, but once they reach this size can grow beyond it much faster by accretion, resulting in a `dip' in the number of SMBHs found around this mass.  Because there is no direct scaling of the
accretion fraction itself, the initial SMBHs formed in this model tend
to be somewhat larger on average than those of Models B and C; indeed,
excluding the unsuccessful Model A, Model D is perhaps the least
promising of all the other models in that it requires the seed
formation efficiency to be $f_{\rm seed} \leq 0.1$ ($ \leq 0.2$ for marginal
success), while the other successful models, although aided by a low
value of $f_{\rm seed}$, can in fact still be made successful with
$f_{\rm seed}=1$.

In Model E, seed SMBHs may only form above a critical redshift $z_{\rm
crit}$.  The table shows that for $z_{\rm crit}=12$ this model proves
successful. Since much
fewer SMBHs are created in this model, the total mass density of SMBHs
at $z=6$ is successfully decreased compared to our original
model. The smaller mass haloes identified at $z\sim6$ are less likely to
have experienced mergers with haloes containing one of these initial SMBHs (they may not even have \emph{any}  progenitors from $z > z_{\rm crit}$ which exceed our resolution limit). Therefore SMBH growth and the
contribution to the total SMBH mass density is preferentially suppressed
at small halo masses and concentrated in the more massive and active systems.  This allows the model to satisfy both of
our constraints simultaneously.
As $z_{\rm crit}$ is decreased, of course, a larger number of initial SMBHs are created and also more smaller mass haloes at $z \sim 6$ will have been able to acquire SMBHs, and so the SMBH mass density increases. The model ceases to be successful for $z_{\rm
crit} < 11.5$ or even marginally successful for $z_{\rm crit} < 8.0$.  Conversely, we expect values of $z_{\rm crit}$ somewhat larger than 12 would allows us to decrease the SMBH mass density even further than shown in Table \ref{tab:summary}.

  It is interesting to note that for the particular value $z_{\rm crit}=12$ the model requires the same value of $f_m$ as Model S (our original simple models with the additional constraints of $f_{\rm seed}=0.1, M_{\rm seed, max}=10^6 \, {\rm M}_{\sun}$), suggesting that the luminous quasars at $z \sim 6$ in that model were all due to SMBHs initially formed at redshifts $z \geq 12$.  This is confirmed by the fact that the number of bright quasars
produced at $z \sim 6$ remains almost constant if we decrease the
value of $z_{\rm crit}$ (keeping $f_m$ the same) but slowly decreases for higher values. 

Model F, in which accretion \emph{and} seed SMBH formation can only take
place in haloes with a sufficiently large virial velocity, can be made
successful with in fact an extremely small SMBH mass density if a
sufficiently large choice of $v_{\rm crit}$ is taken.  The table shows results for a somewhat large value of $v_{\rm crit}=150 \; {\rm km} \, {\rm s}^{-1}$ , but the model remains successful for more conservative values, producing SMBH mass densities more comparable to the other successful models.  We find that $v_{\rm crit} \geq 55 \; {\rm km} \, {\rm
s}^{-1}$ is required for the model to be successful or $v_{\rm crit} \geq 35 \; {\rm km} \, {\rm
s}^{-1}$ for it to be marginally successful. It is interesting to note that marginal
success can be attained with a value of $v_{\rm crit}$ only slightly
higher than the limit already imposed by our mass resolution limit of
$v_{\rm crit}=30\; {\rm km} \, {\rm s}^{-1}$. This reflects the fact
that, as can be seen from the results of Model S in Table \ref{tab:summary}, the reduced value
of $f_{\rm seed}$ and the imposed maximum seed mass are almost enough by themselves
to make the simple model marginally successful. Clearly this model
achieves our goal of reducing the contribution to the SMBH mass
density from the smaller mass SMBHs typically found in small haloes while still allowing SMBHs to form
early in large haloes. However, for the very reason that SMBH
formation takes place only in the largest haloes we find that with our
fiducial value of $f_{\rm seed}=0.1$, nearly all the seed SMBHs formed
would normally have masses larger than our hard limit of $M_{\rm seed,
max} = 10^6 \; {\rm M}_{\sun}$, especially when large values are taken for $v_{\rm crit}$. To avoid having an artificial peak in
the number of SMBHs at this mass we would thus require a significantly
lower formation efficiency ($f_{\rm seed} \sim 10^{-3}$ for $v_{\rm crit}=150 \; {\rm km} \, {\rm s}^{-1}$, although less extreme values of $f_{\rm seed}$ would suffice for more conservative values of $v_{\rm crit}$).

Lastly we see that Model G, in which seed SMBH formation only occurs in
a random fraction of major mergers, also proves successful.  The table
shows results for a low probability of SMBH formation $P_{\rm seed} = 0.02$ but in
fact the model remains successful for probabilities as high as $P_{\rm
seed} = 0.3$ (or $P_{\rm seed} = 0.85$ for marginal success). More
massive haloes will in general experience a larger number of major mergers
over their formation history, and so reducing the overall probability
of seed formation produces seeds preferentially (but not exclusively) in the more massive haloes.  Since the abundance of seed SMBHs is reduced, the early growth of these SMBHs is again likely to become concentrated in the larger haloes, which even if they do not form their own SMBHs are more likely to acquire them through mergers.  Thus much as in Models E \& F the reduced number of seeds combined with a suppression of early growth in smaller haloes allows the model to meet our constraints. Unlike in
Model F, we do not find that the seed SMBH masses tend to exceed $M_{\rm
seed, max}$, presumably because the more stochastic nature of this
recipe means they are not produced exclusively in the most massive haloes.

\begin{figure*}
\includegraphics[width=0.5\textwidth]{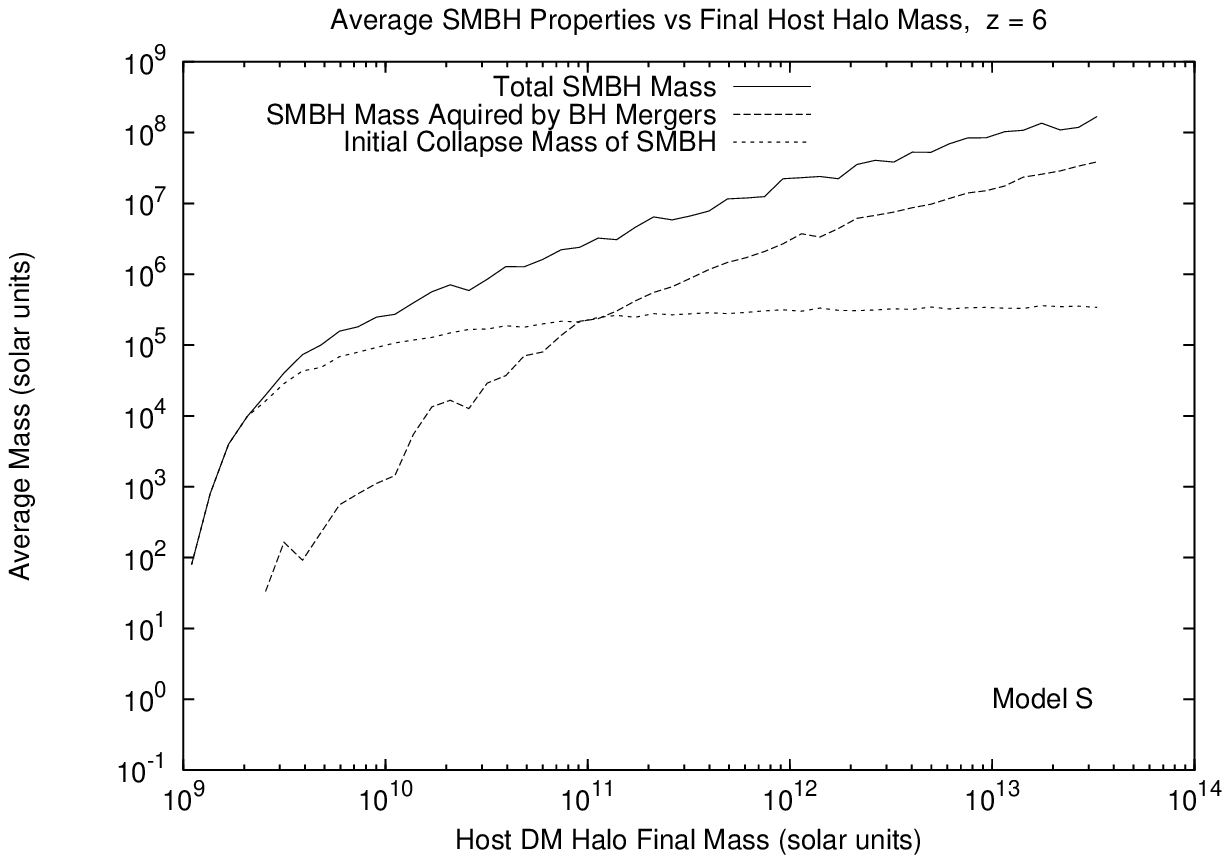}\includegraphics[width=0.5\textwidth]{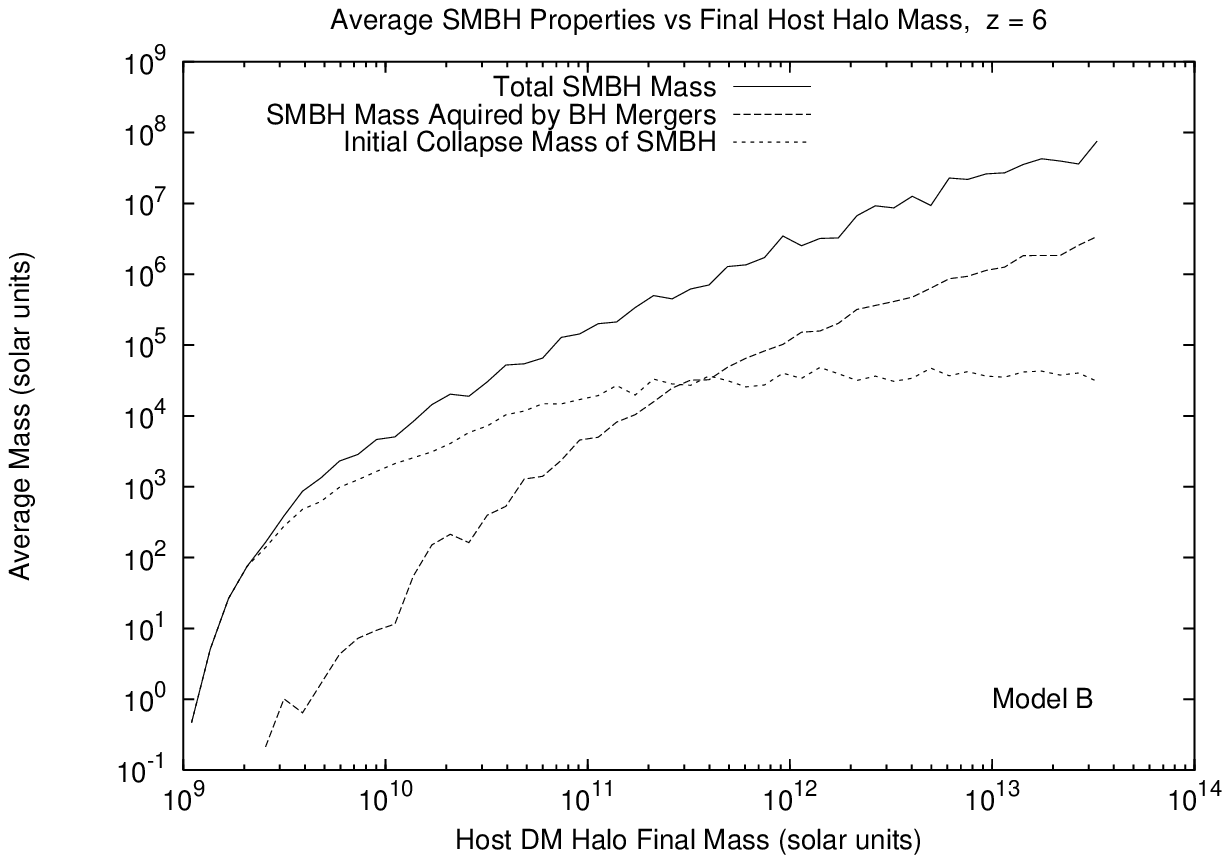} \\ \includegraphics[width=0.5\textwidth]{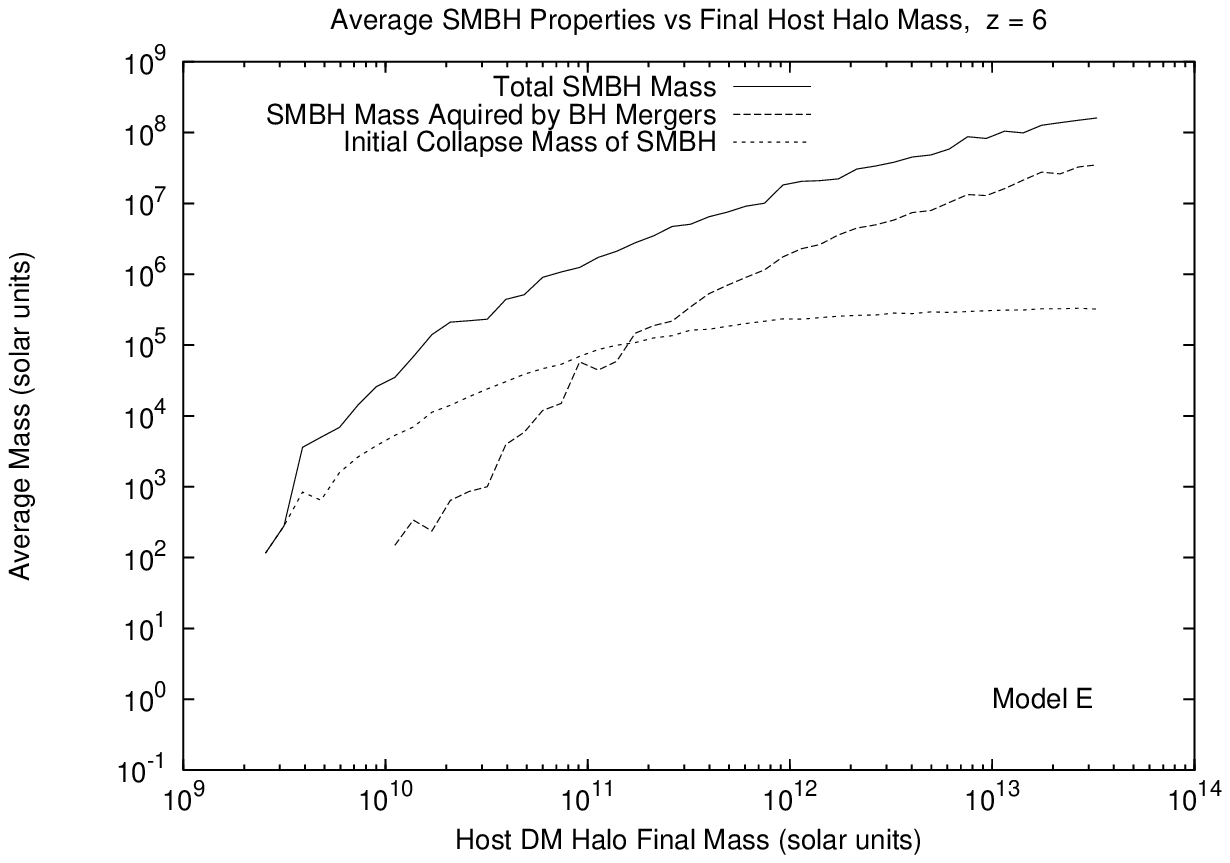}\includegraphics[width=0.5\textwidth]{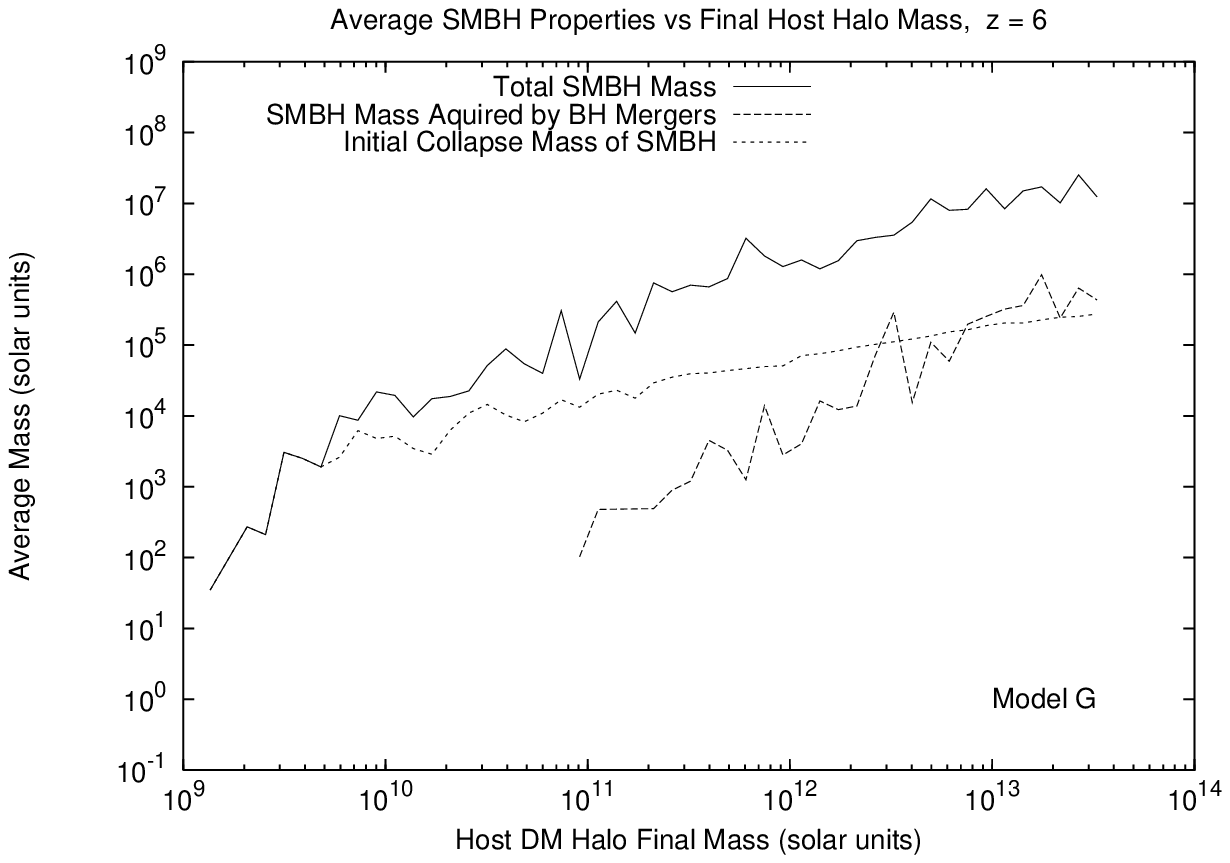} \\
\caption{The mean SMBH mass as a function of host halo mass and the
contribution from black hole mergers and initial collapse (seed SMBHs).
These plots are created by averaging the properties of the SMBH in
final root haloes of different masses at $z \sim 6$ (actually
$z=5.75$) over a large number of realizations, including any
realizations in which no final SMBH is found.  (Note that by
construction the SMBH of a `root halo' always corresponds to the SMBH
in the central galaxy.)  Results are shown for Models S, B, E and G.
 }
\label{fig:goodbad1}
\end{figure*}

In Fig.~\ref{fig:goodbad1} we show the average SMBH mass as a function
of final host halo mass at $z=6$, and the contribution to this from mergers
and the initial collapse (seed).  We show results for three of the
most successful models: B, E and G (using the same parameters as for
the results shown in Table~\ref{tab:summary}) and also for Model S, the adjusted version of our original simple model.  The graph for Model A is not shown, but for $n=1$ would strongly resemble that of Model S.
Comparing the results of Model S in this figure to those of the original simple model (that were shown in Fig.~\ref{fid}d) shows the effects of setting $M_{\rm seed,
max} = 10^6\,{\rm M}_{\sun}$, $f_{\rm seed} = 0.1$.  Notably a larger fraction of SMBH growth in the largest haloes is seen to come from accretion (as opposed to mergers) presumably since many of the SMBHs brought in by satellite galaxy mergers are now smaller because of the reduced initial seed masses.
This effect is even more pronounced in both Models B \& G.
In Model B this is because the scaling of the accretion fraction suppresses SMBH growth in the smaller haloes that
contribute to the SMBH mass via mergers.  While in Model G mergers in fact play a minor role compared even to the
contribution from seeds in all but the largest haloes
since a much smaller number of the merging haloes contain SMBHs, especially at early epochs, than in other models.
This behaviour is mirrored by Model F (not
shown in the graphs) for the same reasons. 
The figure also shows how the fraction of final SMBH mass coming from the
initial seed rises much more slowly with halo mass in Model E than the
other models.  This is because all seed SMBHs are formed at high
redshift, when the host haloes were considerably smaller in mass.
Lastly we note
that the overall mean SMBH mass in the largest haloes is lower for
Model G than in the other models. This is because a sizeable fraction
of massive haloes still do not host a central SMBH at this redshift, lowering the calculated mean.
Once again this applies also to Model F. Note however that this is not
the case in Model E, since all the most massive haloes have managed to form
or acquire SMBHs by $z=6$ in this model.

\begin{figure}
\includegraphics[width=0.5\textwidth]{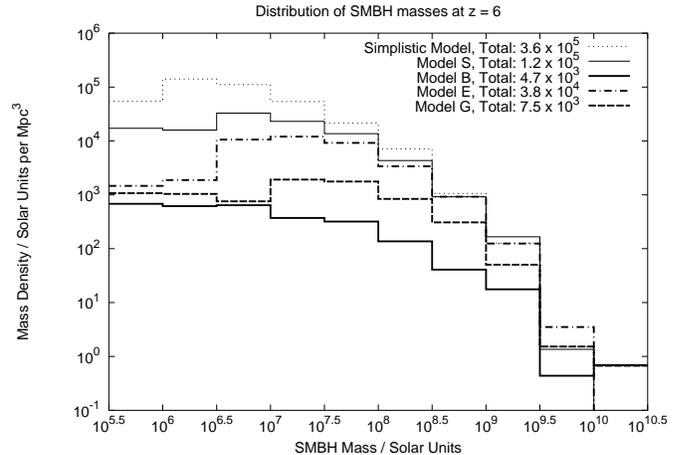} \\
\caption{The contribution to the total SMBH mass density from SMBHs of
different masses (grouped in logarithmic mass bins).  Results are
shown for a Models S, B, E and G and also for our original simple
model.  Note how the mass density in SMBHs larger than $10^9 \; {\rm
M}_{\sun}$ is roughly equal (namely $ \sim 100 \; {\rm M}_{\sun} \,
{\rm Mpc}^{-3}$) for all models shown, because they have all been
normalized to reproduce the observed number density of bright quasars
at $z=6$.  }
\label{fig:goodbad2}
\end{figure}

Fig.~\ref{fig:goodbad2} shows the contribution to the integrated SMBH
mass density from SMBHs of different masses, again for models S, B, E and
G as specified in the table and for the original simple model.  In the
plots for Models B and G we see that the overall reduction in the SMBH
mass density is preferentially at the expense of small mass SMBHs. This
is the hallmark of a successful model, as obviously we must still
produce enough high mass SMBHs to account for the observed luminous
quasars.  We also see how for Model E, the distribution actually turns
over at small masses, because the formation of new, small mass SMBHs
shuts off after the critical redshift. Model S still has
far too much mass in small mass SMBHs, but is nevertheless an improvement over the original model -- the figure shows that the reduced $f_{\rm seed}$ and the presence of $M_{\rm seed,max}$ cause a reduction in the SMBH mass density of $\sim \frac{1}{3}$ at $z=6$, achieved principally by a $\sim 75$ per cent reduction in the contribution from the smaller SMBHs.  The plot for Model A with $n=1$, although not shown here, would in fact strongly resemble that of Model S since its only improvement relative to the
original simple model comes from the reduction in $f_{\rm seed}$ and the
introduction of the limit on maximum seed mass.


Finally, we note that if we remove the cap on seed mass $M_{\rm seed,
max}$ while retaining the same value of $f_{\rm seed}$, although
inevitably some of the seed SMBHs substantially exceed $10^6 \; {\rm
M}_{\sun}$, we find that the number of bright quasars and the SMBH
mass density at $z=6$ is almost unaltered for Models B, C and E and
altered by only 10 -- 20 per cent for Models D and G.  Only for Model
F is the effect very noticeable. Similarly, if we allow the remaining
gas not used in seed formation when $f_{\rm seed}<1$ to be accreted at
the next opportunity (i.e.\ the next timestep), rather than assuming that it is lost, much
(though not all) of the beneficial effects of the reduced seed
formation efficiency are lost. Of the successful Models, D and E are most affected by this and require further reductions in $f_{\rm seed}$ to remain successful.  Additionally Model S can no longer be made successful with 
$f_{\rm seed} \sim 1.8 \times 10^{-4}$.

\section{Comparison with previous work}
\label{sec:others}

Previous models of quasar formation have tended to take one of two
classes of approach for dealing with the initial formation of seed BHs.
One of these has been to simply consider the formation of SMBHs and their
subsequent growth by accretion as a single process, while the other has been to
grow SMBHs from small ($\sim 10 - 100\;{\rm M}_{\sun}$) seed black
holes, generally posited to be remnants of either normal mass or very
massive stars.

When taking the former approach, accretion events which are triggered
in galaxies with no existing SMBH are assumed to form a dense core of
gas which forms the SMBH as the accretion proceeds.  Since the
processes of formation and accretion are not separated, it is not
possible to enforce an Eddington accretion rate, and the luminosities
and light curves are typically taken as a function of the accreted
mass rather than of the SMBH mass.  Examples include the work of
Kauffmann \& Haehnelt (Kauffmann \& Haehnelt 2000; Haehnelt \&
Kauffmann 2000) and Cattaneo, Haehnelt \& Rees (Cattaneo, Haehnelt \&
Rees 1999; Cattaneo 2001). By including the effects of star formation
and feedback these works were able to account for the low to
intermediate redshift behaviour of the quasar population and, by
suitable scaling of the accretion efficiency, to reproduce the local
Magorrian or $M_{\bullet}-\sigma$ relations. Provided that the models
correctly reproduce the abundance of host haloes/galaxies as a
function of morphological type, this then ensures agreement with the
local mass density of SMBHs.  However these papers concentrated on the
low to intermediate redshift objects only, and indeed Cattaneo (2001)
mentions that he is unable to account for the very bright end of the
quasar population on which we have been focusing here.  It is also
worth noting that the latter model was normalized to a somewhat large
value for the local SMBH mass density of
$\rho_{\bullet}(z=0)=10^6\;{\rm M}_{\sun}\,{\rm Mpc}^{-3}$ and did not
track haloes with a virial velocity below $70 \; {\rm km} \, {\rm
s}^{-1}$ -- a rather coarse resolution in light of our results for
Model F in Section~\ref{sec:results}.

In the cases in which SMBHs are grown from stellar mass seeds, a
popular approach is to identify the remnants from the first generation
of stars (the Population III stars) as the seeds.  These would be
formed at high redshift and, due to their lack of metals, be much more
massive than the remnants from later stellar populations.  However the
possibility that SMBH seeds are the remnants from `regular'
star-formation, becoming available once galaxies become large enough
to support a significant stellar population has also been considered
(Haiman \& Loeb 2001).  Volonteri, Haardt \& Madau (2003) recently
used a model based on Population III remnants to successfully match
the observed quasar luminosity function for $1 < z < 4$ and the local
$M_{\bullet}-\sigma$ relation.  They did so without considering the
effects of star-formation, but by assuming that the amount of matter
accreted in mergers scales in the same way as the $M_{\bullet}-\sigma$
relation itself (in fact they used the $M_{\bullet}-v_c$ relation).
Note that there was no \emph{a priori} reason to expect that such an
approach would correctly reproduce the local scaling relation. The
fact that it appears to do so tells us something at a very empirical
level about the role of stellar and/or AGN-driven feedback.
The Volonteri et al. (2003) model is also noteworthy in that it
considered the additional merging time-scale of SMBHs within the host
galaxies, and included a treatment of slingshot ejections. Additionally SMBHs
accreted in minor mergers are tidally stripped of gas and left
`wandering' rather than merging with the central SMBH and so
do not contribute to the overall mass density. Unlike the other
models discussed in this paper, this model also assumed that entropy,
rather than mass, is conserved in SMBH mergers, implying significant
mass loss. Because of these loss mechanisms, the model
effectively had a larger mass budget to work with.

\subsection{A universal BH-galaxy scaling relation}

The tightness of the $M_{\bullet}-\sigma$ relation is so striking that it is 
tempting to try to apply it to galaxies at higher redshift.  However
while it would be interesting to compare the high redshift SMBH mass densities
from our models with the prediction from such a relation, it is unfortunately not a 
completely straightforward matter.

The observed $M_{\bullet} - \sigma$ relation applies principally to the
spheroidal components of galaxies.  Within the confines of a simple 
Press--Schechter model however there is no real way to determine 
the bulge fraction of a given galaxy or deduce its velocity dispersion from
that of the halo and crude approximations must necessarily be made.  
Typically one either simply approximates the bulge velocity dispersion by that
of the entire dark matter halo, or else uses the known relationship between bulge
velocity dispersion and galaxy circular velocity, $v_c$, (Gerhard et al.\ 2001;
Ferrarese 2002) and then approximates the galaxy circular velocity by the virial
velocity of the dark matter halo. 

However such approximations inevitably break down at low redshift, precisely
where one would want to compare them to the observed relation to determine their
effectiveness. This is because at low redshifts the typical mass of a collapsing dark
matter halo corresponds not to a single galaxy but to a group or cluster and it clearly
becomes unfeasible to take the huge virial velocities of such systems as an
approximation to just their centralmost galaxy.

\begin{figure}
\includegraphics[width=0.5\textwidth]{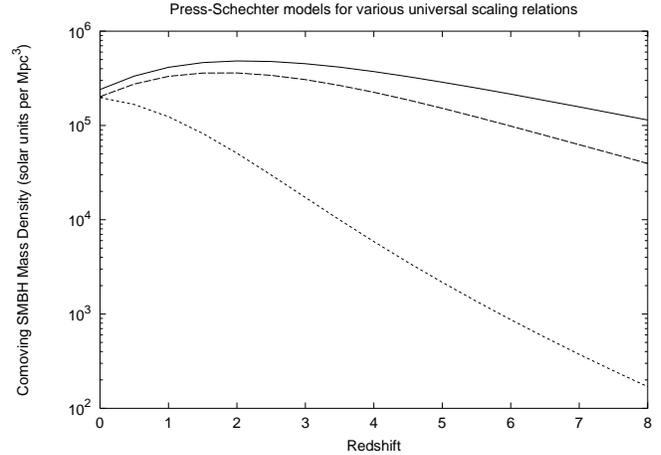} \\
\caption{Predicted SMBH mass densities from simple Press--Schechter
models assuming various scaling relations hold at all redshifts.
\emph{Solid line:} the $M_{\bullet}-\sigma$ relation, assuming that
$\sigma$ can be approximated by the velocity dispersion of the dark
matter halo. \emph{Long-dashed line:} the $M_{\bullet}-v_c$ relation,
assuming that $v_c$ can be approximated by the virial velocity of the
dark matter halo. \emph{Short-dashed line:} the $M_{\bullet}-M_{\rm
DM}$ relation as detailed in the text.  In all cases the initial form of the $M_{\bullet} - \sigma$ and $M_{\bullet} - v_c$ relations is taken to be that of Ferrarese (2002), and a maximum SMBH mass corresponding to $\sigma \sim 350 \; {\rm km} \, {\rm s}^{-1}$  in the relations is applied.}
\label{fig:msigma}
\end{figure}

The best one can do within this simple framework is to `truncate' the approximated
$M_{\bullet} - \sigma$ relation at $\sigma \sim 350 \; {\rm km} \, {\rm s}^{-1}$ (the largest
value found in the observed galaxies on which the relation is derived; Ferrarese 2002;
Merritt \& Ferrarese 2001b), allocating larger systems a SMBH corresponding to this
maximum value.  Using this approach and integrating over different halo masses in the
Press--Schechter distribution (in fact we again use the corrected mass function of Sheth
\& Tormen 1999) produces the mass density curves shown in Fig.\ \ref{fig:msigma}.
We have also investigated using the full merger-tree-based approach to generate such curves,
but the difference is negligible since the contribution from central galaxies still tends to dominate
over that from satellites in haloes with multiple galaxies.

These plots show a rather strange behaviour: the total SMBH mass density does not
increase monotonically, but turns over and starts to decrease at $z
\la 2$ (decreasing to $\sim \frac{1}{2}$ its maximum by $z=0$).  This cannot be avoided
without raising our cut-off in $\sigma$ to unfeasibly high values. Whether
this behaviour (which would seem difficult to explain physically) implies that the
$M_{\bullet}-\sigma$ relation is not invariant with time (as suggested by the work of Di Matteo et al.\ 2003) or is simply a consequence of the crude
approximations used to relate the properties of the galactic spheroid to the dark matter halo is unclear, 
but it shows that results from such approximations need to be interpreted with some caution.  Thus although the curves show a higher SMBH mass density (by a factor of $\sim 2 - 4$) at $z \sim 6$ than that which we advocated
as necessary for success in our own models, we do not feel we can draw any strong conclusions from this given the clear uncertainties.

A more sophisticated investigation into the possibility of a redshift independent $M_{\bullet}-v_c$ scaling has been presented by Wyithe \& Loeb (2002a, 2003).  Rather than take the exact form of the local $M_{\bullet}-v_c$ relation, they consider simply that there is some scaling $M_{\bullet} \propto v_c^{\gamma}$ independent of redshift.  They approximate $v_c$ by $v_{\rm vir}$, but do not track the bright end of the quasar population below $z \sim 2$ to avoid the worst of the problems discussed above.  In their latest model (Wyithe \& Loeb 2003) quasar activity is triggered by major mergers and $\gamma = 5$, a value which could arise naturally if quasar fuelling is self-regulated (Silk \& Rees 1998), although earlier models (Wyithe \& Loeb 2002a) have also shown a fair degree of success for other values.  The initial formation of SMBHs is not explicitly treated, although a minimum host halo mass is imposed, and accretion is assumed to take place at the Eddington luminosity over the dynamical time-scale of the galactic disc.  Due to the nature of the model it is not possible to explicitly ensure that the accreted mass inferred from this luminous period exactly balances that enforced by the scaling relation at each step, but nevertheless the model is remarkably successful.   It is one of the few other models to tackle the reproduction of the bright SDSS observations at $z > 5.8$, which it manages while still matching the observed lower redshift luminosity functions.  It will be interesting to see in the future whether merger-tree-based models where individual SMBH masses can be directly tracked are able to reproduce such successes.

Lastly we briefly consider an alternative suggestion (Ferrarese 2002; see also Haiman \& Loeb 1998) that the local $M_{\bullet}-\sigma$
relation in fact reflects an underlying relation between SMBH mass and dark matter
halo \emph{mass} that remains constant in time.  We demonstrate the effects of
such a scenario within a Press--Schechter model with the short-dashed curve in Fig.\ \ref{fig:msigma}.  To do this we have again approximated the galaxy circular velocity by the halo virial velocity
and used the spherical top-hat model to relate the virial mass of a halo ($M_{\rm DM}$) to its virial velocity (see Somerville \& Primack 1999) to obtain the zero redshift relation\footnote{Ferrarese (2002) derives a similar
relation, but has a different value for the coefficient. This is due
to her use of a fitting formula from Bullock et al.\ (2001) which
featured a misprint in its original publication (J. S. Bullock,
private communication).}:
\begin{equation}
\frac{M_{\bullet}}{10^8\, {\rm M}_{\sun}} \sim 0.015\left(\frac{M_{\rm
DM}}{10^{12}\, {\rm M}_{\sun}}\right)^{1.82}
\end{equation}
We then assumed that the relation is redshift invariant and as before imposed a maximum SMBH mass corresponding to the largest systems in the observed $M_{\bullet} - \sigma$ relation, so as to minimize the low redshift problems related to group and cluster haloes.  As seen,
this predicts a dramatically different scaling of SMBH mass density
with redshift.  This is
due to the fact that the relation between virial mass and velocity is different for
haloes that collapse at different redshifts (haloes that collapse at
higher redshift are denser, and so rotate faster for a given mass).
The SMBH mass density now grows monotonically with time,
but so steeply that the predicted total mass budget at $z=6$ is only
$\rho_{\bullet}(z \sim 6)=870 \; {\rm M}_{\sun} \, {\rm Mpc}^{-3}$. 
Even given the uncertainties involved in these approximations, this still makes such a universal relation between SMBH and halo mass seem unlikely.
We
have already seen how much difficulty hierarchical models experience
in producing enough luminous quasars at these redshifts without
overproducing the much higher mass budget we had adopted. Of the
models considered, only Model F with a rather extreme minimum virial
velocity for seed SMBH formation of $v_{\rm crit}=150\; {\rm km} \, {\rm
s}^{-1}$ had such a low integrated SMBH mass density.  
Thus even if the $M_{\bullet}-M_{\rm DM}$ relation 
cannot be completely ruled out, it would likely comprise an even more
stringent constraint on models than the ones we have considered.

\section{Conclusions}
\label{sec:conc}

We have investigated a basic scenario for the formation and fuelling of
SMBHs within hierarchical dark matter merger trees. In this scenario, SMBH
formation and fuelling are triggered by major mergers, and once
`activated' SMBHs accrete at their Eddington rate until their fuel is
consumed.  We have not attempted to model the depletion of fuel
arising from star formation and related feedback processes, but
instead have concentrated upon meeting the very high redshift
constraints on the quasar population as a minimum criterion for
success.  To do this we have combined the number density of observed
high redshift quasars at $z \sim 6$ with an estimated limit on the
high redshift `SMBH mass budget' inferred from local measurements of
the SMBH mass density and the observed quasar luminosity function at
$z\la3$.  In our simplest model, a constant fraction of
the gas in the halo falls to the centre after every major merger where it forms a seed SMBH if there is no pre-existing SMBH, or else accretes on to the existing SMBH. We
have shown that such a model, when the single free parameter (the
accreted fraction of gas $f_m$) is normalized to reproduce the number density
of high redshift quasars, badly exceeds the SMBH mass budget at
$z=6$ placing too much mass in smaller SMBHs. This is a direct result of the steep slope of the dark matter halo mass
function in CDM models, and echoes similar problems with reproducing
the galaxy luminosity function is such models (e.g. White \& Frenk
1991). 

We further showed that the seed SMBHs formed in this scenario can tend to unphysically large values in large mass haloes. We
found that we could counter this by imposing a reduced efficiency of seed formation, in which
only about ten per cent of the gas freed by the major merger goes into forming a SMBH (the rest being lost back to the galaxy), and a
maximum seed mass of $10^6 \; {\rm M}_{\sun}$. We also found reducing the seed
formation efficiency to much smaller values $\sim 10^{-4}$ allowed us
to bring the total SMBH mass density down to a reasonable level, but
at the expense of probably overpredicting the number of lower redshift
quasars.  Though within the current framework of our model such a low efficiency would seem perhaps hard to justify physically.

Including now a reduced seed formation efficiency of $f_{\rm seed}=0.1$ and upper mass
limit $M_{\rm seed,max}=10^6 \; {\rm M}_{\sun}$, we then presented a broad range of models involving simple modifications to the
the formation of seed SMBHs or to their growth via accretion.  These
modifications were designed to bias SMBH growth toward high mass
haloes, reducing the relative overabundance of small SMBHs seen in the original model and thus allowing us to reproduce the observed number density of
luminous quasars without exceeding the imposed mass budget. We have
shown that several of these recipes do indeed prove successful; our conclusion would not be changed by relaxing the maximum seed mass to $10^7 \; {\rm M}_{\sun}$, indeed most of the successful variants are affected only slightly even by a complete removal of the limit. Our findings for
these model variants are summarized below:

\begin{enumerate}

\item A model in which the fraction of baryons accreted by the SMBH
following a merger, $f_m$, was scaled as $(1+z)^n$ was not
successful. We suspect that any non-contrived scaling with redshift
will not solve the problem at hand.

\item Introducing a strong scaling of $f_m$ with halo virial velocity
was very successful. We found very good results with $f_m \propto
v_{\rm vir}^2$ and slightly inferior, but still acceptable, results
with the relation suggested by Kauffmann \& Haehnelt (2000) and given
in Eqn.\ \ref{eqn:kh}. The latter was advocated by these authors in
order to reproduce the local $M_{\bullet}-\sigma$-type scaling
relations.

\item Decreasing the number of seed SMBHs able to form, either
through a redshift cut-off, a virial velocity cut-off or a purely
stochastic effect, also had the desired effect. Of these, we feel that
the model with the velocity cut-off was the least preferable as it
tends to produce only very large mass seeds.

\item Allowing accretion to take place only when SMBHs exceed $10^6 \;
{\rm M}_{\sun}$ can also allow our constraints to be met, but is
perhaps less effective than the other models. Its success is dependent upon the reduced seed formation efficiency (i.e.\ it requires $f_{\rm seed} \leq 0.1$).

\end{enumerate}

The mechanisms adopted in the successful models identified here may all
be associated with physical processes.  For example, if reionization
can prevent gas from collapsing in haloes as large as $55 \; {\rm km}
\, {\rm s}^{-1}$, which is not implausible, this would be sufficient to make our models successful, as seen in Model F. As we have noted before, our conclusions will likely also be pertinent to other possible BH formation scenarios.  In particular if the dominant mechanism
for forming seed BHs is core collapse of massive $\sim 200 \; {\rm M}_{\sun}$
Pop III stars, or even of supermassive ($\sim 10^6 \; {\rm M}_{\sun}$) stars in early metal-poor dwarf galaxies (Bromm \& Loeb 2003), then it is quite reasonable to expect that this
mechanism would shut off at a redshift around $z_{\rm crit} \sim
10-12$, as the IGM becomes polluted with metals and proto-stellar gas
clouds are more likely to fragment into stars with a `normal' IMF
(e.g. Bromm \& Clarke 2002; Bromm \& Loeb 2003). This scenario is similar to our successful
Model E. The strong scaling of accretion fraction with halo circular
velocity seen in Models B and C could arise from stellar or AGN-driven
feedback (e.g. Silk \& Rees 1998). While a stochastic element to seed SMBH
formation or fuelling in major mergers, as in Model G, may be motivated
if we consider that the efficiency of merger-triggered inflows on
larger (kpc) scales seems to depend on many variables such as impact
parameter, orbital inclination, and host galaxy morphology (e.g. Mihos
\& Hernquist 1994).

Lastly, we have drawn attention to the difficulties faced when trying to assume a redshift invariant `universal
$M_{\bullet}-\sigma$ relation' in the context of these simple Press--Schechter based models; and also we have shown that if a
relation between SMBH mass and halo virial mass is taken to be redshift
invariant, then the inferred total SMBH mass density at $z\sim6$ is so
low that producing enough luminous quasars within this budget will be
extremely difficult in any CDM model.

While the proliferation of parameters and modelling uncertainties
needed to attempt a full self-consistent treatment of galaxy and
quasar formation is somewhat daunting, this work suggests that it is
necessary if we are to understand the important population of quasars
and AGN and how they are related to galaxies. We hope to use the
insights gained here to build a more physically motivated, joint model
of quasar and galaxy formation in the near future. 

\section*{Acknowledgments}

We thank Martin Haehnelt, Paul Hewett, Avi Loeb, and Priyamvada
Natarajan for useful discussions.  We should also like to thank the anonymous referee for helpful comments and suggestions. JMB acknowledges the support and
funding of a PPARC studentship. RSS acknowledges support from a PPARC
theory rolling grant during early stages of this work, and hospitality
at the IoA through the visitor's program during later stages. RSS is
currently supported by AURA through NAS5-26555.  ACF thanks the Royal Society for support.

\end{document}